\documentclass[preprint,10pt]{aastex}
\usepackage{emulateapj5}
\newcommand{\eten}[1]{\mbox{$10^{#1}$}}

\newcommand{\degree}{\mbox{$^{\circ}$}}

\newcommand{\as}{\mbox{\arcsec}}

\newcommand{\kms}{\mbox{km s$^{-1}$}}
\newcommand\cmv{\mbox{cm$^{-3}$}}

\def\lsim {$\rlap{\raise.4ex\hbox{$<$}}\lower.55ex\hbox{$\sim$}\,$}

\newcommand\submm{submillimeter}

\newcommand\fir{far-infrared}

\newcommand\nir{near-infrared}
\newcommand\uv{ultraviolet}

\newcommand{\lsun}{\mbox{L$_\odot$}}
\newcommand{\msun}{\mbox{M$_\odot$}}

\newcommand{\lbol}{\mbox{$L_{bol}$}} 
\newcommand{\tbol}{\mbox{$T_{bol}$}} 

\newcommand{\mean}[1]{\mbox{$\langle#1\rangle$}} 
\newcommand{\rinf}{\mbox{$r_{inf}$}} 
\newcommand{\lsmm}{\mbox{$L_{smm}$}} 
 
\newcommand{\form}{H$_2$CO}

\newcommand{\hcopi}{H$^{13}$CO$^+$}

\newcommand{\nthp}{N$_2$H$^+$}
 
\input{epsf}

\newcommand{\Snu}{\mbox{$S_{\nu}$}}

\newcommand{\Inunorm}{\mbox{$I_{\nu}^{norm}(b)$}}
\newcommand{\Snut}{\mbox{$S_{\nu}(\theta_{mb})$}}
\newcommand{\kappanu}{\mbox{$\kappa_{\nu}$}}
\newcommand{\Td}{\mbox{$T_{d}$}}
\newcommand{\Tdr}{\mbox{$T_{d}(r)$}}

\newcommand{\ppc}{pre-protostellar core}
\newcommand{\ppcs}{pre-protostellar cores}

\newcommand{\beam}{\mbox{$\theta_{mb}$}}

\newcommand{\chisq}{\mbox{$\chi_r^2$}}
\newcommand{\chisqsed}{\mbox{$\chi_{SED}^2$}}

\newcommand{\sisrf}{\mbox{$s_{ISRF}$}}
\newcommand{\lint}{\mbox{$L_{int}$}}
\newcommand{\aeff}{\mbox{$a_{eff}$}}
\newcommand{\ro}{\mbox{$r_{o}$}}
\newcommand{\ri}{\mbox{$r_{i}$}}
\newcommand{\Go}{\mbox{$G_{0}$}}
\newcommand{\lobs}{\mbox{$L_{obs}$}} 

\begin{document}

               
\title {\bf Tracing the Mass during Low-Mass Star Formation. 
III. Models of the Submillimeter Dust Continuum Emission 
from Class 0 Protostars}
\author {Yancy L. Shirley and Neal J. Evans II}
\affil{Department of Astronomy, The University of Texas at Austin,
       Austin, Texas 78712--1083}
\affil{yshirley@astro.as.utexas.edu}
\affil{nje@astro.as.utexas.edu}
\and
\author{Jonathan M. C. Rawlings}
\affil{Department of Physics and Astronomy, University College London,
        Gower Street, London WC1E 6BT}
\affil{jcr@star.ucl.ac.uk}

 
\begin{abstract}

Seven Class 0 sources mapped with SCUBA at 850 and 450 \micron\
are modeled using a one dimensional radiative transfer code.  The
modeling takes into account heating from an internal protostar,
heating from the ISRF, realistic beam
effects, and chopping to model the normalized intensity profile
and spectral energy distribution.  Power law density models,
$n(r) \propto r^{-p}$, fit all of the sources; best fit values are
mostly $p = 1.8 \pm 0.1$, but two sources with aspherical emission
contours have lower values ($p \sim 1.1$). Including all sources,
$\mean{p} = 1.63 \pm 0.33$.
Based on studies of the sensitivity of the best-fit $p$ to variations 
in other input parameters, uncertainties in $p$ for an envelope model are
$\Delta p = \pm 0.2$.
If an unresolved source (e.g., a disk) contributes $70\%$ of the flux
at the peak, $p$ is lowered in this extreme case and $\Delta p = ^{+0.2}_{-0.6}$.
The models allow a determination of the
internal luminosity ($\mean{L_{int}}  = 4.0$ \lsun ) of the central 
protostar as well as a characteristic dust temperature for mass
determination ($\mean{T_{iso}} = 13.8 \pm 2.4$ K).
We find that heating from the ISRF strongly affects 
the shape of the dust temperature profile and
the normalized intensity profile, but does not
contribute strongly to the overall bolometric luminosity of Class 0
sources.  There is little evidence for variation in the
dust opacity as a function of distance from the central source.
The data are well-fitted by dust opacities for coagulated
dust grains with ice mantles (Ossenkopf \& Henning 1994).
The density profile from an inside-out collapse model (Shu 1977)
does not fit the data well, unless the infall radius is set so
small as to make the density nearly a power-law.

\end{abstract}



\section{Introduction}

Modern theories of star formation predict the evolution of the 
density structure, $n(\vec{r},t)$, and velocity structure, 
$\vec{v}(\vec{r},t)$, of the 
gas and dust envelope of protostars.  
Optically thin dust emission at submillimeter wavelengths provides an
observational constraint on the density distribution of the protostellar 
envelope and therefore constrains theoretical models of star formation.

Class 0 protostars represent an early, highly embedded phase
during the formation of low mass stars ($M < $ few $\msun $).  
The original evolutionary sequence for low mass protostars 
(Class I, II, and III) is based on the shape of the 
spectral energy distribution (SED) from observations at near and mid-infrared 
wavelengths (Lada 1987). Low mass protostars are thought to evolve from a thick
dusty envelope where most of the energy is re-radiated by dust in the far-infrared (Class I)
to progressively less embedded objects with near and mid-infrared excesses from dusty disks 
(Class II or classical T Tauri stars and Class III or weak-line T Tauri
stars). The discovery of extremely embedded objects with submillimeter
telescopes led to a new class of protostars, 
Class 0 objects, which are so highly enshrouded that their near infrared 
emission has generally not been detected and their SEDs peak longward of 100 \micron .
Observationally, Class 0 sources are
cores that have $\lbol / \lsmm \leq 200$, where \lsmm\ is the total
luminosity detected longward of 350 \micron\ (Andr\'e et al. 1993).   
Alternatively, Class 0 sources are
characterized by $\tbol \leq 70$K, where \tbol\ is the temperature of a blackbody
with the same mean frequency as the observed SED (Chen et al. 1995). 
Based on the relative numbers of Class 0 and Class I objects,
it has been argued that the timescale for Class 0 objects is short, 
perhaps $10^4$ years (Andr\'e et al. 2000).  Class 0 objects have
powerful outflows, which suggest high accretion rates (Bontemps et al. 1996).  
As data improve, some Class I sources are being reclassified to
Class 0 sources (e.g., Shirley et al. 2000, Young et al. 2001), 
suggesting a reexamination of this argument (see also Visser, Richer, \&
Chandler 2001).

It is important to understand the structure of the envelope of Class 0 objects 
since they are likely to be the earliest observed phase of star formation 
with a central accreting protostar (Andr\'e et al. 2000).
The emission from an optically thin dust shell observed 
at an impact parameter, $b$, with dust opacity, $\kappa _ {\nu}$, 
that does not vary with radius, is given by
\begin{equation}
I_{\nu}(b) = 2 \kappa _ {\nu} \int_{b}^{r_o} B_{\nu}(T_{d}(r)) 
\rho (r) \frac{r}{\sqrt{r^{2} - b^{2}}} dr 
\end{equation}
(Adams 1991), where $r_o$ is the outer radius.
For an optically thin
envelope (at all wavelengths) dominated by a central source of luminosity, $L_{int}$,
the dust temperature distribution can be approximated by a power law:
\begin{equation} 
\Tdr \propto \left( \frac{L_{int}}{ r^2 } 
      \right)^{\frac{1}{4 + \beta }}   
	\; \; \propto \; L_{int}^{q/2}r^{-q} ,
\end{equation}
where $q = 2/(4 + \beta )$
(cf. Doty \& Leung 1994) and $\beta$ is the power law exponent of the
dust opacity ($\kappa \propto \nu ^{\beta}$), which typically
lies between 1 and 2 in the submillimeter.  
If we also assume the density distribution follows a power law, 
$n(r) \propto r^{-p}$, then in the Rayleigh-Jeans limit ($\Td \gg h\nu /k$),
the specific intensity integral simplifies to $I_{\nu} \propto b^{-m}$, 
where $m = p + q - 1$. Previous studies of the
density structure of Class 0 objects assumed a temperature power
law of the form $\Tdr \propto r^{-q}$ (Walker et al. 1990, Ladd et al. 1991,
Chandler \& Richer 2000,  Hogerheijde \& Sandell 2000, 
Shirley et al. 2000, and Motte \& Andr\'e 2001).
However, this approach is $\bf{not}$ valid in the outer
envelopes of low luminosity sources due to a breakdown 
in the Rayleigh-Jeans approximation at wavelengths shorter than 1 mm and
due to external heating from the interstellar radiation field (ISRF) 
(Shirley et al. 2000).  
It then becomes necessary to calculate \Tdr\ self consistently 
in the integral in Equation (1) to
reveal the density distribution of the envelope.

In Paper I (Shirley et al. 2000), 21 low mass cores  within 325 pc
were observed at 850 \micron\ and 450 \micron\
using the Submillimeter Common User Bolometer Array (SCUBA) (Holland et al. 1999)
on the JCMT 15-m radio telescope. 
Thirteen sources from the original sample were classified as Class 0, 
three sources having been previously classified as Class I.
In this paper, we present 1D dust radiative transfer models for seven of the 
Class 0 sources selected for spatial isolation
and high signal-to-noise radial profiles ($\left< S/N \right>_{peak} = 60$): 
B335, B228, L723, IRAS03282+3035, L1448C, L1527, and L483.  
B335, B228, IRAS03282+3035, and L1448C appear circular in the dust continuum 
maps down to the 20\% contour, while L1527, L483, and L723 are clearly
aspherical at that level.  The bolometric temperature
ranges from 23K (IRAS03282+3035) to 52K (L483) and the bolometric luminosity
ranges from 1.2 \lsun\ (IRAS03282+3035 and B228) to 13 \lsun\ (L483).
In Paper II (Evans et al. 2001), three \ppc s (L1512, L1544, and L1689B)
were modeled using one-dimensional radiative transfer and a beam convolution code.
In this paper, the seven Class 0 sources will be modeled using the
same procedure used in Paper II, with the addition of an
internal luminosity source.  The modeling procedure and inputs are
discussed in Section 2.  
We use B335 as a test object (Section 3.1) to model the sensitivity to
the model input parameters.  
Individual sources are modeled in Sections 3 and 4, while the 
implications of our best fit models are discussed in Section 5.

\section{1D Dust Radiative Transfer Modeling}

	The 1D dust models are calculated from a modified version of the 
Egan, Leung, and Spagna (1984) continuum radiative transfer code. This code 
iteratively calculates the equilibrium dust temperature on a 1D radial grid 
by simultaneously solving the 
combined moment radiative transport equations in quasi-diffusion
form and the energy balance equations as a two-point boundary value problem.  
The radiation field is constructed by solving a set of 
ray equations along impact parameters, $b$, through the cloud. 
There are four physical inputs to the radiative transfer code: the density 
distribution, $n(r)$; the internal source luminosity, $L_{int}$; the scaling
factor for the interstellar radiation field, $s_{ISRF}$; 
and the dust opacity, $\kappa _{\nu}$.
In addition, the radial grid (100 points) and impact parameters, as well as 
the wavelength grid (59 wavelengths) are chosen to cover the relevant range 
so that results are insensitive to the details of these grids.
The equilibrium dust temperature distribution, \Tdr , which is the output
from the dust code, is used by an observation simulation code
(Paper II) to calculate the normalized
radial intensity profiles, \Inunorm , 
by solving Equation (1), performing beam convolutions, and simulating chopping.
The code also models the observed SED, \Snut , by convolving the model
intensity with the beam (\beam) used in each observation.  

The internal radiation field is assumed to be a blackbody with 
an effective temperature of 6000 K.  Since all of the objects we are modeling
are very opaque at near-infrared wavelengths, 
the exact shape of the spectrum of the internal
radiation field is not important; only the total internal luminosity
is important.  The shape of the ISRF was determined from COBE 
results (Black 1994) plus the cosmic microwave
background (CMB) and the \uv\ component of the ISRF ($\lambda \leq 0.36 
\micron$) from van Dishoeck (1988).  This ISRF is considerably stronger in the
infrared than the previous versions of the ISRF (Mathis, Mezger, \&
Panagia 1983) (see Figure 2 of Paper II for a plot of the different ISRFs).
We modify the strength of the ISRF by multiplying all portions of the
ISRF spectrum except the CMB with a factor denoted by \sisrf .  
For the models discussed in this paper, we used a coarse grid 
with $\sisrf = 0.3$, $1.0$, and $3.0$.  These factors for the far-UV (FUV)
portion of the ISRF correspond to 0.45\Go ,
1.5\Go , and 4.5\Go\ respectively, where \Go\ is integrated FUV flux between 
$91.2$ nm and $220$ nm in units of $1.6 \times 10^6$ erg s$^{-1}$ cm$^{-2}$ 
(cf. Hollenbach et al. 1991).  In Paper II, the best fit to 
the observed radial profiles and SED suggest a lower strength to 
the ISRF ($\sisrf \sim 0.3 - 0.5$).

The dust opacity, $\kappa _{\nu}$, was taken from the models of Ossenkopf 
and Henning (1994) for grains that have coagulated for \eten5 yr at 
a density of \eten6 \cmv, both with (OH5) and without (OH2) 
accreted ice mantles.  We assume a gas to dust ratio of 100 by mass to 
convert OH5 opacities (per gram of dust) to opacity per gram of gas.
OH5 opacities have been successful
at reproducing the SED of both high mass (van der Tak et al. 2000) and low
mass star forming cores (Paper II).  
OH5 opacities at submillimeter wavelengths can be 
approximated by a power law ($\kappanu \propto \nu ^\beta$) 
with $\beta \sim 1.85$ and an opacity of
$1.8 \times 10^{-2}$ cm$^2$ per gram of gas at 850 \micron.
The OH2 dust opacities are lower at \fir\ wavelengths
but up to 1.6 times higher at submillimeter and millimeter wavelengths
than OH5 dust opacities. The cross-over point is near 350 \micron\ 
(see Fig. 2 of Paper II).

The JCMT beams derived from AFGL618 and Uranus
were used for the beam convolution of the model profiles at 850 and 450 
\micron\ (see Figure 1).
It is very important to convolve the model specific intensity distribution
with a realistic beam profile to derive an accurate estimate of the
envelope density distribution.  Using a gaussian beam shape instead of the
actual beam profile can profoundly distort the interpretation of
the power law index, $p$, by up to 0.5 (Shirley et al. 2000). 
Since the beam shape was stable during the second half of the night in
April 1998 (see Paper I), nine Uranus profiles were added together
to use an average beam profile with high signal-to-noise out to 70\as\ 
from the center of the map.
No planets were available to make beam maps for the January 1998 
sources (L1527, IRAS03282, L1448C); so the secondary calibrator, AFGL618 was observed.  
The January 1998 observations were taken in the first
half of the night when the beam shape is not so stable and changes
shape continuously as the telescope cools.  
AFGL618 is much weaker than Uranus at
850 and 450 \micron ; therefore the beam maps have much lower signal-to-noise
at large radii.  The January beam map was produced by averaging together three
AFGL618 maps observed during the same time of night that our objects were 
observed.  Because the 450 \micron\ profile of AFGL618 becomes too noisy 
beyond 34\as, the average Uranus
beam profile from April was used for the 450 \micron\ January beam beyond
34\as .  These beam calibration difficulties introduce larger 
uncertainties, $\Delta p \sim 0.1$, in the best fit models for the January sources.

Many of the radial profiles from Paper I show a turn-down
in the intensity profile beyond 60\as\ from the center of the map.  
This turn-down could be due to a steepening of the density profile or 
due to the effects of chopping.  To test the
possibility of a steeper density profile, we must account for the chop throw. 
The SCUBA observations were chopped in azimuth with a 120\as\ chop throw.  
Since the SCUBA array is 2D, all of the positions within a single annulus chop
different distances from the center of the map.  Therefore, our 1D model can only
include an approximate simulation of the true effects of chopping. 
The details of how we simulate chopping has a noticeable effect on the 
shape of the model intensity profiles beyond 60\as ; 
consequently, we do not attempt to model normalized intensity profiles
beyond this radius.

The agreement between the model and the data can be quantified in terms
of the reduced chi-squared for the normalized intensity profile at wavelength
$\lambda$:
\begin{equation}
\chi^2_{\lambda} \; = \; \sum_{i}   \left[  \frac{ \left( 
I_{\nu}^{norm}(b_i) \right)_{obs} - \left( I_{\nu}^{norm}(b_i) \right)_{mod}  }
                            { \sigma_{I}(b_i)}  
             \right]^2 / \; N_{b} \; ,
\end{equation}
where 
$I^{norm}_{\nu}(b_i)$ is the azimuthally averaged, normalized intensity
in a circular aperture at impact parameter $b_i$, 
$\sigma_{I}(b_i)$ is the uncertainty in the data, and $N_b$ is the number of 
impact parameters. Only points spaced by a full beam are used in computing 
$\chi^2_{\lambda}$
to avoid introducing correlations.  We calculate $\chi^2_{\lambda}$ for the
850 and 450 \micron\ profiles and define \chisq\ as the 
sum of $\chi^2_{850}$ and $\chi^2_{450}$.
The signal-to-noise was higher for the 850 \micron\ maps and therefore
the $\chi^2_{850}$ has the most weight in determining the best fit. 

The agreement between the model SED and the observed one is quantified by 
$\chi _{SED}^2$, calculated from a similar equation:
\begin{equation}
\chi^2_{SED}  = \; \sum_{i} \left[  \frac{ S_{\nu_i}^{obs} - S_{\nu_i}^{mod} }
                          { \sigma_S(\nu_i) }
             \right]^2 / \; N_{\nu} \; ,
\end{equation}
where $S_{\nu_i}^{obs}$ is the observed flux into a beam and $S_{\nu_i}^{mod}$
is the modeled flux into the same beam. When photometry at the same wavelength
with different beams exists, the points are both considered in calculating
$\chi^2_{SED}$. Wavelengths shorter than 60 \micron\ are 
not included because they are expected to be optically
thick and therefore very sensitive to asymmetric geometries (e.g., outflow
cavities, flattened envelopes), which we are currently unable to model.
In addition, much of the far-infrared photometry has large and uncertain
calibration errorbars, and the opacities as a function of frequency are
uncertain. For all these reasons, poor fits to the SED are not considered a
serious problem in constraining the density distribution; however we
often find that the best \chisq, which considers only the profiles, occurs
for the model with the best $\chi^2_{tot} = \chisq + \chi_{SED}^2$.

In fact, the SED and the normalized radial intensity profiles 
provide nearly orthogonal constraints on model parameters (\S 3).  
The SED is sensitive to the strength of the ISRF and a mass$\times$opacity 
product (Paper II).  In particular, the flux density at 850 \micron\
constrains the mass, while the full SED provides information on the
the variation of dust opacity with frequency, subject to the caveats
mentioned above. Measurements of the flux density into different beams
at the same wavelength provides some constraint on the density distribution,
but the {\it shape} of the density distribution is much better constrained
by the normalized radial profile (\Inunorm).

The simplest models to test are single power laws,
\begin{equation} 
n(r) = n_f \left( \frac{r}{r_f} \right) ^{-p} \;\; ; \;\; r \in [r_i,r_o] 
\end{equation}
in \cmv\ for the gas density.  The density, $n_f$, is normalized to
a fiducial radius, $r_f$, of 1000 AU and represents the total gas number
density ($n = n(\rm{H}_2) + n(\rm{He}) + ... = \rho/(\mu m_{\rm{H}})$, $\mu = 2.29$).  
There are seven parameters in the power law models: the power law 
exponent, $p$; the
density at a fiducial radius, $n_f$; the inner radius, $r_i$; the outer
radius, $r_o$; the internal source luminosity, $L_{int}$; 
the dust opacity, $\kappa _\nu$ 
and the strength of the interstellar radiation field, $\sisrf$.  
In these models, the shape of the density profile, defined by $p$
is constrained by \Inunorm, and only slightly affected by other parameters.
In contrast, $n_f$ and hence the mass, are constrained by the observed
flux density at optically thin wavelengths; we use $S_{\nu}(850)$ because
the calibration errors are lowest. The resulting mass depends inversely
on the opacity at 850 \micron\ ($\kappanu (850)$) and weakly on other
parameters. 

We also test models of inside out collapse (Shu 1977), hereafter referred
to as Shu77 models. These are characterized by seven parameters: $a_{eff}$,
the effective sound speed; the radius that encloses the infalling gas,
\rinf ; $r_i$; $r_o$; $L_{int}$; $\kappa _\nu$; and \sisrf . 
Inside \rinf, the density distribution tends toward $n(r) \propto
r^{-1.5}$, while $n(r) \propto r^{-2}$ outside \rinf. The shape of the
density profile is thus set by \rinf, while the normalization is set by
\aeff. We fix \aeff\ based on observations of optically thin spectral lines; 
thus Shu77 models have no freedom in the normalization of the density
profile, unlike power law models.

The constraints on the other parameters are similar for the two types
of density profiles.
The internal source luminosity (\lint) is constrained by the integral
of the SED and secondarily by \sisrf. 
The internal luminosity dominates the bolometric luminosity over heating 
from the ISRF for sources with $\lint \geq 1 \lsun$.
The model internal luminosity is tuned until the model bolometric luminosity
($L_{bol}^{mod}$) matches the observed bolometric luminosity
($L_{obs}$, Paper I), using the same method to integrate 
over the SED.  The model bolometric luminosity is calculated using the same
wavelengths and beams as the observed bolometric luminosity.
$L_{obs}$ and $L_{bol}^{mod}$ may be greater than $L_{int}$ for low
luminosity sources because the ISRF adds energy.
In some cases, especially for more luminous sources \lobs\ underestimates
\lint\ because the beams used at some wavelengths do not capture all the
emission (see Butner et al. 1990 for a full discussion).
The choice of the overall
opacity law (\kappanu ($\nu$)) is constrained by the {\it shape} of
the SED once other parameters, like $p$, \lint, and \sisrf\ are constrained.

Models of B335 are used to test our assumptions of nearly 
orthogonal constraints and the effects of changing various parameters 
in the model (\S 3).

\section{Testing Model Parameters -- B335}

B335 (IRAS 19345+0727, L663) is an extensively studied Class 0 object 
within the Barnard 335 dark cloud.  
Its submillimeter emission is very nearly circularly symmetric 
(Huard et al. 1999, Paper I, 
Motte et al. 2001).  The core is the one of the best cases for
a collapse candidate as deduced from models of observed asymmetric line
profiles in CS and \form\ (Zhou et al. 1993, Choi et al. 1995).
Rotation, if any, is very slow (Frerking et al. 1987, Zhou, 1995), making
simple spherical models reasonable.  
B335 has an outflow that lies nearly in the plane of the sky along an 
east-west direction
(Goldsmith et al. 1984).  There is little direct evidence in the submillimeter
continuum maps of extensions or flattening along or perpendicular
to the outflow direction,
making this core a suitable choice for 1D modeling.
Harvey et al. (2001) have recently studied B335 using a near-infrared 
extinction mapping technique to probe the density structure
(cf. Alves, Lada, \& Lada 2001).  These observations provide
an important check on the consistency of our submillimeter continuum 
models. We shall model the density structure of the outer envelope of B335
using both a single power law and the inside-out collapse model of Shu (1977).

\subsection{Power Law Models}

A $p=2$ power law is a good starting model because it has the same density 
distribution as a singular isothermal sphere, until it is truncated at an 
inner radius.
The outer radius was chosen to allow proper simulation of chopping
at the distance of B335 (250 pc; Tomita et al. 1979).
The chop throw for our observations was 120\as, corresponding to 30,000 AU at 
the distance of B335.  Both the radial profile and the observed 
beam profile contain useful information 60\as\ from the center of the map; 
therefore, to allow the beam to chop onto the model density distribution for
radial points 60\as\ from the center, the modelled density distribution must 
extend to twice the chop throw (240\as ) or 60,000 AU at the distance of B335. 
NICMOS observations of B335 indicate that reddening of background stars 
blend into the background noise beyond 125\arcsec\ (Harvey et al. 2001),
which is within the range of the outer radii tested in our models.
We initially chose the inner radius such that $r_o/r_i = 1000$
so that $r_i = 60$AU for B335.  For the initial test model, we used 
OH5 opacities. The fiducial density, $n_f$, was varied to match the 
observed flux at 850 microns; $S_{850}^{obs} = 3.91 \pm 0.22$, and the 
internal luminosity, $L_{int}$, was varied until the model \lbol\
matched $L_{obs} = 3.1 \pm 0.1 \lsun$ 
(Paper I). The resulting best fit was $n_f = 1.7\times 10^6 \cmv$ and 
$L_{int} = 2.5 \lsun$ (see Table 1). This basic model was used to 
test the effects of varying $r_o$, $r_i$, $\kappa _\nu$, $p$, \sisrf , and
distance.  The sensitivity of $p$ and $L_{int}$ to changes in the other parameters
is summarized in Table 2.

Changes in $r_o$ result in changes in the total envelope mass proportional to 
$r_o^{3-p}$ for a power law; however, the column density along a line of 
sight, and hence the emission at long wavelengths into a fixed beam 
becomes insensitive to $r_o$ for $p>1$ 
($N \propto r_i^{1-p}(1-(r_i/r_o)^{p-1}$).  
For the test model, the density is
$4.7 \times 10^2$ \cmv\ at 60,000 AU.  This low value  
is probably not realistic as the dense core is embedded in an extended 
cloud (Frerking et al. 1987).
Even if a constant density of $10^3$ \cmv\ is used for the lower limit to the
density in the envelope, doubling the outer radius 
to 120,000 AU only increases A$_V$ by slightly less than 1 magnitude of visual 
extinction. 
This lower limit is used for all subsequent models. 
Allowing the density to fall below $10^3$ \cmv\ results in no 
significant change in the best fit fiducial density or internal luminosity. 
Factor of two changes in \ro\ have negligible effect on the best fit
parameters. If \ro\ is made small enough ($\ro \leq 15000$ AU),
the fit degrades (Table 1), but this $r_o$ is smaller than the extent of
observed emission in the SCUBA map.

While the column density increases as $r_i$ decreases, beam dilution
is more important.
The SCUBA beams are much larger than the inner radii
in our models (the 450 \micron\ beamwidth of 8\as\ corresponds to a
radius of 520 AU for our nearest source, B228); therefore 
changes in the inner radius will not have substantial effects unless
the inner radius gets too large.
Factor of two changes in \ri\ have negligible effects; a factor of 4
increase in \ri\ begins to degrade the fit to
the 60 \micron\ point on the SED (Figure 2).
However, 60 \micron\ observations can be affected
by deviations from spherical geometry, so the 60 \micron\ flux should not 
be considered a strong constraint on one dimensional models.  
We do not model the effects of a disk (see Section 5.5). 

The result of these tests is that the best-fit model parameters are not 
sensitive to changes of factors of 2 in \ro\ or \ri. Larger changes
worsen the fit, indicating that there is no evidence in our data for
either inner or outer boundaries in the power laws.
	
Calculations of dust opacities differ substantially between dust models,
especially at long wavelengths; for example OH5 and OH2 dust models
have opacities at $\lambda = 1000$ \micron\ that are 6 and 12 times
those of the model for dust in the diffuse interstellar medium (Draine
\& Lee 1984). With no further information, these differences
would result in substantial uncertainties in $n_f$, or equivalently the
mass since $S_{\nu} \propto M \kappa _{\nu}$. However, the SED is
strongly affected by the choice of opacity law and can be used to 
distinguish dust models (e.g., Butner et al. 1991, van der Tak et al.
2000) if the density distribution is independently constrained.
If OH2 opacities are used instead of OH5 opacities, 
the best fit power law remains $p=2.0$ with $n_f$ about one-third that
of the test model.  The higher \kappanu (850) of OH2 dust 
requires less column density to match the flux at 850 \micron .  
None of the power law fits using OH2 dust produce enough emission to match 
the SED at $100$ \micron\ $\geq \lambda \geq 350$ \micron\ (Figure 2), 
yielding quite large values of $\chi_{SED}^2$.  
Changing the submillimeter opacity by a factor of a few does not 
strongly affect the {\it shape} of the model density profiles, confirming
our conclusions about the orthogonality of the radial profile and the SED.  
A dust opacity that varies with radius could in principle affect the fit 
to the shape of the density profile; 
however, we see no evidence for variation of opacity with
radius (see \S 5.3).

Other models of star formation predict density distributions different from
the singular isothermal sphere (e.g. McLaughlin \& Pudritz 1997, Ciolek \& 
Basu 2000). Without explaining the details of these models we 
note that the differences often can be approximated by 
variations in the power law index ($p$) in the bulk of the envelope.
We therefore considered models with $p=1.5$, $2.0$, and $2.5$ (Figure 3).  
Both the $p=1.5$ and $p=2.5$ models fit very poorly (Fig. 3).
The $p=2.5$ model shows that the predicted radial profile becomes
very sensitive to the beam shape for steep power laws.
Variations in $p$ of 0.1 produce noticeable changes in the intensity
profiles. 
In summary, we
find that the $p=2.0$ initial test model is the best fit profile for OH5 dust 
and $\sisrf = 1.0$, and that variations in $p$ of 0.1 can be distinguished
if all other parameters are fixed.

Heating by the ISRF affects the temperature distribution in the
outer envelope.  The temperature 
levels off and eventually rises again towards the outer boundary of the 
cloud. The location of the minimum in the temperature is 
dependent on the ISRF strength and \lint\ (Figure 4). 
The strength of the ISRF could affect the intensity profiles and the SED.
We tested these effects on the best fit power law by multiplying 
the \uv\ to \fir\ portion of the ISRF by factors of 
$\sisrf = 3$ and $\sisrf = 0.3$.
With $\sisrf = 3.0$, the best fit is obtained with $p=2.2$ and $L_{int} = 1.5$ \lsun.
Obviously, the greater contribution to the flux from the ISRF explains
why a lower internal luminosity is appropriate. Even though $\chi_r^2$ and
$\chi_{SED}^2$ are not significantly different from the standard model, the 
predicted SED is too bright at FIR wavelengths and severely underestimates 
the 60 \micron\ flux.   
Using $\sisrf = 0.3$, the best fit is obtained with $p=1.8$ and 
$L_{int}=3.3 \lsun$ (Figure 4 \& 5). This model clearly yields the best 
fit to the intensity profiles of all of the power law models that 
we have considered, but the model SED has too much flux at shorter wavelengths 
(i.e., 60 \micron ). 
A lower strength of the ISRF is consistent with the best fit models in
Paper II.  A study of [CII] and [OI] emission lines with ISO towards
B335 is consistent with a slightly higher strength of the ISRF, $2$ -- $3$\Go\ 
(Nisini et al. 1999).  Clearly, there is uncertainty in the strength of the
ISRF around B335.
We conclude that uncertainties of a factor of 10 in \sisrf\
cause changes in $p$ by $\pm 0.2$, but that the SED provides some
constraints that can decrease this uncertainty.

The effect of the uncertainty in distance was checked on the initial 
test power law by assuming B335 was at
half the published distance (125 pc; cf. Harvey et al. 2001).  
Since the core is now closer, smaller inner and outer radii (30 pc and 
30,000 pc) were used.
The best fit power law index and shape of the model profile are
very insensitive to the uncertainty in the distance (\chisq\ increases by
less than $3\% $). 
The best fit $\lint = 0.7$ \lsun, and the total mass in the model
decreased by a factor of 2.
The mass of a power law envelope scales as $M_{env} \propto n_f r_o^{3-p}$.
For factor of 2 uncertainties in the distance, \lint\ will be uncertain
by factors of about 4, the mass by factors of 2 (for $p=2$), but
the best fit $p$ is unaffected.

The effect of changing the observed beam shape 
was tested by substituting the January 1998 beamshape
into the best fit power law model for B335.  
The difference in beamshape between January
and April 1998 beams was greater than the difference between
individual beam profiles observed during April 1998.  The best fit
power law was slightly lower, $p = 1.7$, indicating that uncertainties
in the beam shape may effect the interpretation of the best fit
model by as much as $\Delta p \sim 0.1$.

A $p=1.8$ power law with $\sisrf = 0.3$ provides the best fit to
the radial profiles for B335 (Figure 5).
However, the \chisqsed\ is not as good for this model as for the test 
model. A somewhat different opacity law would fit better.
The fit to the radial profiles is not perfect, with some systematic
deviations as a function of impact parameter that suggest deviations
from a simple power law, but given uncertainties in the beam
shape, we consider this fit to be adequate.

In summary, our tests support the idea that the normalized radial profile 
(\Inunorm) and the SED provide largely orthogonal constraints. 
\Inunorm\ constrains the shape ($p$) of the density profile, while
\Snu(850) constrains the normalizing density, $n_f$, and hence the mass.
The density and mass enter as a product with the opacity at 850 \micron,
leading to uncertainties in the mass equal to the uncertainties in the
opacities. However, the rest of the SED depends on the overall opacity law;
within the limitations of spherical models, the overall SED constrains
the opacity law and hence reduces the uncertainty in mass.
Variations in other parameters, such as opacity, distance, and \lint,
have negligible effect on $p$ (Table 2). Variations in the ISRF have the largest
effects: a factor of 3 in either direction in \sisrf\ 
cause variations in $p$ of $\pm 0.2$.
Factors of 3 in \sisrf\ can be constrained by the SED, so larger
errors are unlikely.  Opacity and the strength of the ISRF strongly
affect $L_{int}$ (Table 2).

\subsection{Shu77 Inside-Out Collapse Models}

The Shu77 density distribution has successfully matched asymmetric line
profiles seen towards Class 0 sources including B335
(Zhou et al. 1993, Choi et al. 1995, Hogerheijde et al. 2000).  
Based on Monte Carlo radiative transfer models of CS and \form\ asymmetric 
line profiles, 
Choi et al. found that a Shu77 model with $r_{inf} = 6200$ AU and 
$a_{eff} = 0.23$ \kms\ was the best fit for an inside-out collapse model 
for B335.
We mostly constrain $a_{eff}$ to this value, unlike the procedure used by
Hogerheijde \& Sandell (2000), who allowed \aeff\ to be a free parameter.
This lack of flexibility in the normalizing
density puts Shu77 models at a disadvantage compared to the power law models with
two free parameters.

The Shu77 density distribution with $r_{infall} = 6200$ AU was tested
assuming $\sisrf = 1.0$, and OH5 opacities. Compared to the best-fit
power law model, the density in this Shu77 model is about 5 times less.
There is simply not enough material to match the SED: \lint\
had to be increased to
$6.5$ \lsun\ to match $L_{obs}$ because the
lower optical depths allowed too much radiation at $\lambda < 60$ \micron,
which is not observed.  Even with the large \lint,
the fits to both the SED and radial profiles are extremely poor (Figure 6).  
The model radial profiles are too flat.  This is not surprising since a 
$p=1.5$ power law model did not fit the observed radial profiles.

Shu77 models would need  smaller infall radii to preserve an
$r^{-2}$ density distribution throughout a greater extent of the envelope.
Smaller infall radii provide better fits to the radial profile.
The best of these models, with $r_{inf} = 1000$ AU and $\sisrf = 0.3$,
has a very good \chisq.
However, $r_{inf} = 1000$ AU is less than the FWHM of the SCUBA beam 
at 450 \micron.
Thus, the best fit Shu77 model actually resembles a $p=2$ power law in
all portions of the profile except the central beam.
A small infall radius is highly unfavored by models of profiles of
molecular lines ($\chi ^2$ is 20 times worse, Choi et al 1995).  
Further radiative transfer modeling of the 
molecular line emission using the best fit density profiles from the
dust models and considering possible abundance variations in the
molecular tracers may resolve this discrepancy.

Thus, the Choi et al. model has two problems: the density is too low
to match the observed SED, and the density distribution is not steep
enough to match \Inunorm.
Because other opacity models combined with Shu77 models have fit the
SED of B335 (Zhou et al. 1990), the first problem is not necessarily fatal. 
The bigger problem is the failure to match \Inunorm.

The near infrared extinction study of Harvey et al. (2001) found that
the Choi et al. (1995) parameters for a  Shu77 model did fit the density 
structure of the outer envelope, but require a scaling of the density by a
factor of 5.  Such a large increase would require a higher \aeff\
or B335 to be much closer.  A higher \aeff\ is not supported by the 
narrow linewidth observed towards B335 (e.g., Mardones et al. 1997). 
As shown in the power law models
of B335, a closer distance does not strongly affect the interpretation of
the best fit radial profiles; therefore, an infall radius of $6200$ AU 
would still not fit the observed normalized intensity profile.
This solution, which worked for Harvey et al.,
does not work for our data.  A small infall radius would be required
to match the SCUBA profiles with a Shu77 model.  
The increased density required by Harvey et al. (2001) to model their
\nir\ extinction measurements makes their densities agree well with
our best-fit power law model, which is based on OH5 opacities at
850 \micron. The OH5 opacity at 850 \micron\ and the extinction used
by Harvey et al. in the \nir\ thus produce consistent density estimates.
This agreement supports the validity
of OH5 dust, as does our comparison of dust and virial masses (\S 5.2).

\section{Sources}

The other sources were modeled with less extensive exploration of
parameter space, guided by the results of the B335 models.

\subsection{B228}

B228 is a deeply embedded protostar (IRAS 15398-3359) 
which appears very similar to B335 at 850 and 450 \micron .  
The B228 contours are nearly circularly symmetric 
with extended envelope emission and a high
signal-to-noise profile.  Unlike B335, B228 is much less studied and
therefore has less published SED information (see Paper I). B228 is
an excellent candidate for 1D dust modeling of the envelope emission
and should be the subject of further detailed study.  
We shall try both power law and Shu77 models.

A range of power law indices were tried with noticeable changes 
in the \chisq\ with changes of $\Delta p = 0.1$. 
The best fit power law is $p=1.9$ with a lower ISRF ($\sisrf = 0.3$) and
an internal luminosity of 1.0 \lsun\ (see Figure 7).
Since B228 is closer than B335 (130 pc, see Paper I), the outer radius was decreased 
to 30,000 AU.  While the $p=1.9$ power law is a reasonable fit to both the
profile and SED, it has the same problem as the B335 power law fits.
The model profile is too steep in the inner portion of the envelope and
too shallow in the outer portion of the profile.  The deviations 
may be due to subtle changes in the beam shape or to departures from a
single power law.  Far-infrared photometry with better spatial
resolution and more wavelength points would help to
constrain the SED.

The best fit Shu77 collapse model has an infall radius of $1000$ AU
and an effective sound speed of $0.23$ \kms .  
The linewidth of optically thin lines in B228 
(e.g.,  $\Delta v$(N$_2$H$^+$) in Mardones et al 1997)
is similar to that of lines in B335.
Therefore an $a_{eff} = 0.23$ \kms\ is reasonable.
As with B335, the infall radius is
within the central SCUBA beam.  The best fit Shu77 model was able to
simultaneously match the observed bolometric luminosity and the flux
at 850 \micron .  The fit to the profiles are not as good as the fit
of the best fit power law, and $\chi^2_{SED}$ is considerably worse.

\subsection{L723}

The \submm\ emission from L723 (IRAS 19156+1906) has asymmetries only
in the lowest contours of the 850 and 450 \micron\ emission.  The higher
contours are very circularly symmetric. A quadrupolar
outflow has been detected, indicating that L723 may contain a close binary 
(Palacios \& Eiroa, 1999) within a common envelope.  Alternatively,
the outflow may be due to a single source with a very large opening
angle (Hirano et al. 1998).  The largest extensions in the
lowest contours of the SCUBA maps are roughly aligned with
an east-west outflow.  The unique characteristics of the outflow structure
and the possibility that L723 may be a proto-binary system
warrant a detailed study of the dust continuum structure of the envelope.

	L723 is best fitted by a $p=1.8$ power law with $\sisrf = 0.3$ (Figure 8).  
The fit to the profiles and SED is very good using OH5 dust.  The model
internal luminosity is $2.6$ \lsun\ indicating a contribution of $0.5$ \lsun\
to the bolometric luminosity from the ISRF.  The large contribution from
the ISRF may be due to the large size of
model ($60,000$ AU) and the greater distance ($300$ pc; Goldsmith et al. 1984)
of L723.  Emission was detected beyond 65\as\ ($\sim 20,000$ AU) 
in the 850\micron\ SCUBA map, larger than the outer radius quoted
by Motte \& Andr\'e (2001) of $14,000$ AU from 1.3 mm continuum maps.  As
seen in the test models of B335, decreasing the outer radius to $30,000$
AU has a negligible effect on the shape of the radial profile and
the interpretation of the best fit power law.

	The best fit power law is very similar to the other circularly
symmetric cores B335 and B228.  While L723 may be a proto-binary system, there
is no strong evidence for differences in the one dimensional dust models
of L723 and B335 or B228.  Multiple dimensional dust models are
needed to probe the extended structure at low contour levels in the
SCUBA maps.  The envelope structure of L723 is complicated by
the extensive observed outflows.  The impact the outflow structure has
on the overall distribution of material in the envelope cannot
be effectively modeled with only one spatial dimension.

	The best fit Shu77 model has a small infall radius ($1000$ AU)
and effective sound speed of $0.29$ \kms .  The fit to the profile and
SED is very good.  As with B335 and B228, the infall radius is within the central 
450 \micron\ SCUBA beam.  There is no evidence for a break in the power
law density distribution in the outer envelope beyond the central beam.

\subsection{IRAS03282+3035}

IRAS 03282+3035 is a nearly circular symmetric core with slight 
asymmetries in the lowest contour of the 850 and 450 \micron\ emission. 
There is a well studied outflow 
(e.g., Bachiller et al. 1994) that is nearly perpendicular to the 
extended submillimeter continuum emission.  
The signal-to-noise of the radial profiles was lower for this object
(Paper I).

The best fit to IRAS 03282+3035 is a $p=1.9$ power law with 
lower strength of the ISRF, $\sisrf = 0.3$ (see Figure 9).    
The model fits the 850 \micron\ profile marginally better than the 
450 \micron\ profile. 
Since IRAS 03282+3035 was observed during January 1998,
the beam profile is much more uncertain, resulting in larger uncertainties
in the best fit model.  $L_{int} = 1.0$ \lsun\ for the best fit model,
accounting for nearly all of the bolometric luminosity.

We continue to use a distance of $220$ pc (\v{C}ernis 1990) for
consistency with Paper I, but the
recent study of de Zeeuw et al. (1999) finds a greater distance of 
$318$ pc using Hipparcos parallaxes of the Perseus OB association.  
At the larger distance,
the internal luminosity and mass in the envelope would increase
by a factor of two.  As was seen in the models of B335, the best
fit power law index is not sensitive to distance and remains
$p=1.9$.

	The best fit Shu77 model again has a infall radius of $1000$ AU 
within the central SCUBA beam.  Molecular line observations towards
IRAS03282+3035 indicate linewidths that are similar to B228 and B335
($\Delta v($N$_2$H$^+) = 0.49$ \kms , Mardones et al. 1997); however,  a 
higher effective sound speed ($0.26$ \kms ) than used for B228
is needed to match the 850 \micron\ flux and $L_{bol}$ simultaneously using OH5 dust.

\subsection{L1448C}

	L1448C (also called L1448-mm) is a well studied Class 0 protostar in the 
vicinity of 3 other deeply embedded protostars in the Perseus molecular cloud
(L1448N(A \& B), L1448NW; see O'Linger et al. 1999, Barsony et al. 1998).
L1448C drives a powerful, highly collimated outflow (Bachiller et al. 1990)
for which proper motion has been detected in extremely high velocity SiO maps
(Girat \& Acord 2001).  H$_2$O maser emission has been detected (Chernin 1995)
indicating the presence of very dense gas.  Recently, the infrared spectrum
from 6\micron\ to 190\micron\ has been observed with ISO and extensively studied
(Giannini et al. 2001, Nisini et al. 1999). As for IRAS03282+3035, we continue
to use a distance of 220 pc, but 318 pc is more likely (de Zeeuw et al. 1999);
the effects on the model parameters are described in the previous section.

The submillimeter contours are circularly symmetric with a weak bridge of 
emission extending between L1448C and L1448N/L1448NW (Barsony et al. 1998, 
Chandler \& Richer 2000, Paper I).
L1448C is located 82\as\ from the nearest source (L1448N); therefore
we shall attempt to model L1448C as an isolated source with the strong caveat 
that the outer portion of the observed radial profile (i.e., $> 45$\as ) is
contaminated with emission from L1448N and L1448NW.  Only points in the 
radial profile $\leq 45$\as\ from L1448C are modeled.

	The best fit power law was $p = 1.7$ with \sisrf\ $= 1.0$ (Figure 10).  
An outer radius of $45,000$ AU was used for the best fit model.
This model cannot be compared directly to observations beyond half the distance
between L1448C and L1448N/L1448NW ($9000$ AU projected on the sky).  
However, the large outer radius is appropriate for the model since all of the sources 
in the SCUBA map are clearly embedded in a diffuse envelope that extends to the 
edge of the map.  The model of the 450 \micron\ profile falls off too 
steeply for the best-fit 850 \micron\ profile, which could be caused by
a larger \sisrf\ or uncertainties in the beam.

	This power law density structure is slightly flatter than 
that of the other protostars with
symmetric submillimeter emission.  If the entire profile of L1448C is contaminated
from emission from L1448N/L1448NW, then the observed intensity profile would be
flatter than for an isolated source resulting in a flatter best fit $p$.  However,
it seems unlikely that the northern sources could significantly alter the conclusions 
of the modeling of the entire profile since they are located at least $18000$ AU away.

This is the only modeled source for which $\sisrf = 1$  gave the best fit.  
This source is also the only source modeled 
with multiple cores observed within the SCUBA
map.  It is possible that this indicates that the ISRF is indeed greater in the 
vicinity of L1448C; however, the overall change in the $\chi ^2_{tot}$ between 
$\sisrf = 0.3$ and $\sisrf = 1.0$ is only $25\%$.

\subsection{L1527}

L1527 is a well studied Class 0 source (IRAS 04368+2557) in the Taurus
molecular cloud complex characterized by asymmetric submillimeter and
far-infrared emission extending 
from the southeast to northwest  direction (Ladd et al. 1991, Chandler \& Richer 2000, 
Hogerheijde \& Sandell 2000, Paper I, Motte \& Andr\'e 2001), 
strong evidence for rotation (Goodman et al. 1993, Zhou et al. 1996, Ohashi et al 1997), 
a molecular outflow in the east-west direction (MacLeod et al. 1994, Bontemps et al. 1996), 
and an associated near-infrared nebula (Eiroa et al. 1994).  The normalized radial
intensity profile was not well fitted by a single power law in Paper I,
as it shows considerable curvature.
Previous studies of the dust continuum emission (Hogerheijde \& Sandell 2000, 
Motte \& Andr\'e 2001) have estimated the density power law to be near 
$p = 1$, much lower than the best fit power laws of B335 and B228.

The best fit power law model (see Figure 11) is a shallow
$p=1.1$ power law with $\sisrf = 0.3$, in agreement
with previous dust continuum studies of L1527.  
The fit to the 850 \micron\ profile is within the errorbars except for the last 
point.
The observed profile is slightly flatter than the model $p=1.1$ power law
between 5000 and 6500 AU.  The 450 \micron\ model profile is flatter than the
observed profile in the inner portion but matches extremely well in the 
outer envelope. 
These discrepancies may be explained by the inability of the one-dimensional dust code
to appropriately model the observed large asymmetries in the dust continuum 
maps 
or may be explained by the problematic January 1998 beam profiles (cf. Section 2).
The beam effects cannot be strong enough to change our conclusion that the density structure
in L1527 is clearly flatter than the density structure of B335 or B228.  
The curvature in \Inunorm\ can be matched by simple
power laws when \Tdr\ and the effects of the beam and chopping are properly
simulated.  
The fit to the SED is poor at far-infrared wavelength as indicated by the the 
poor $\chi^2_{SED}$ for all of the models in Table 1.

The sensitivity of the model to asymmetries
can be tested by eliminating sectors in the azimuthally averaged intensity
profile.  L1527 is elongated in a south-east to north-west direction.  If a
symmetrical sector of 70\degree\ centered along the major axis is removed from
the azimuthal average, then the best fit power law index changes only 
to $p = 1.0$.  
Therefore, the density distribution appears to fall off more slowly even
perpendicular to the extension; thus the lower value of $p$ appears to be
real and not an artifact of azimuthally averaging an elongated intensity
map. Uncertainties on the order of $\Delta p \sim \pm 0.1$ result from 
azimuthally averaging the L1527 continuum emission.  The uncertainty in
best fit $p$ is consistent with effects seen in 1.3mm continuum maps of 
L1527 (Motte \& Andr\'e 2001).

A  standard Shu collapse model was not tested since the density distribution only
flattens to $p \sim 1.1$ in a small region around the infall radius and
is much steeper everywhere else in the envelope.
Zhou et al. (1996) were able to model molecular line
emission with a Terebey-Shu-Cassen model (TSC, Terebey et al. 1984),
a perturbation of the Shu collapse model to account for rotation.
The one-dimensional averaged TSC profile is similar to
$r^{-1.1}$ in the region inside the infall radius (5400 AU).
A more detailed understanding of the density structure requires higher dimensional
radiative transfer models to effectively model the asymmetric, 
flattened structure due to rotation and outflow cavities.

\subsection{L483}

L483 (IRAS 18148-0440) is another example of a Class 0 protostar with 
large asymmetries observed in the submillimeter continuum maps.  The 850 and 450 \micron\
emission is extended in the northeast to southwest direction (Paper I), 
nearly perpendicular to the observed outflow direction (Parker 1988).  Water maser
emission has been detected towards L483 (Xiang \& Turner 1992) as well as 
an associated   
near-infrared nebula (Hodapp 1994).  Molecular line studies identify this source 
as a possible collapse candidate (Mardones et al. 1997) but with several 
confusing signatures due to infall and outflow (Park et al. 2001). 

The dust continuum emission from L483 is best fitted by a shallow ($p=1.2$)
power law with $\sisrf =0.3$ (Figure 12).  The model SED and 850 \micron\
radial profile match very well but the 450 \micron\ model profile is not steep 
enough in the outer half of the envelope.  
This may be partially due to uncertainties in the beam
profile at 450 \micron , a variation in the dust opacity in the outer
envelope, or a variation in the temperature structure due to geometrical effects
compared to a one dimensional model.  
Once again, OH5 dust matches the observed SED well;
however, the 160 \micron\ point from Ladd et al. (1991) is much higher than the 
surrounding 100 \micron\ and 190 \micron\ points, 
perhaps indicating a larger calibration error 
than the published value.  The internal luminosity of the model is  
$13.0$ \lsun .  This is the most luminous source in this sample of Class 0
objects, making the contribution to the bolometric luminosity from the 
ISRF negligible compared to the internal luminosity, but the effects on the
temperature profile are still substantial.
The resulting temperature profile flattens near the outer regions probed by
SCUBA observations making a single temperature power law a poor approximation.  
Our model is consistent with $1.3$ mm continuum models of Motte \& Andr\'e
(2001), which assume a power law temperature distribution, $\Tdr \propto r^{-q}$,
($p=1.2 \pm 0.6$ for $q = 0.2 \mp 0.2$).
The best fit model becomes optically thick around 30 \micron\ due to
the flatter power law and higher luminosity internal source.  
   
	Like L1527, the L483 profiles were not well fitted by a
single power law in Paper I, but a power law density distribution
does fit well when the beam, chopping, and \Tdr\ are correctly modeled.
The best fit power law is close to $p = 1$, for both these sources with
elongated dust emission contours.
Both L1527 and L483 have \nir\ nebulae
associated with the core.  However, both cores are not entirely
similar since the extensions seen in the submillimeter continuum contours 
are in opposite directions with respect to the outflow axes.  
Geometrical projection effects are clearly important for these two sources.    
Tafalla at al. (2000) and Pezzuto et al. (2001) suggest this
source is in the transitional phase between Class 0 and Class I sources
based on the observed properties of its outflow and far-infrared colors.
While still a Class 0, it does have the largest \tbol\ (52K) of the
sources we model here.
While the envelope structure is clearly different from the
more symmetrical cores B335 and B228, it is not clear whether the
lower $p$ indicates a more evolved state or an intially less
spherical envelope.  
Higher dimensional dust modeling is required to fully understand 
the density structure of the outer envelope.

	No Shu collapse models were tested since the power law model
was flatter than the Shu collapse model.  However, it is possible that
a higher dimensional model that includes effects of rotation (TSC) may fit
the intensity profile.

\section{Discussion}

\subsection{Density Distributions}

	The properties of the best fit power law for all of
the Class 0 sources are listed in Table 4.
Power law models successfully fit the outer ($r > 1000$ AU) 
envelope structures of Class 0 protostars.  
Five of the Class 0 sources are well fitted by steep power laws
($p = 1.7 - 1.9$, B335, B228, L723, IRAS03282+3035, L1448C),
while the two cores with elongated emission contours 
are fitted by shallow power laws ($p = 1.1 - 1.2$, L1527, L483).  
We quantify the degree of elongation in Table 4 by giving the aspect
ratio (ratio of long axis to short axis) of the 20\% contour.
L1527 and L483 have the largest aspect ratios.
Both L1527 and L483 also have \nir\ nebulae,
suggestive of aspherical density distributions.
The fiducial 
densities for all of the Class 0 cores are consistent with the central
densities ($n_f \sim n_c \sim 10^6$ cm$^{-3}$) of the denser \ppcs\ modeled
as Bonnor-Ebert spheres in Paper II.

	The best fit Shu77 models for B335, B228, L723, and IRAS03282+3035
have small infall radii (within the central SCUBA beam).  The Shu77 models
look very similar to the best fit power laws, but generally do not fit as
well. There is no strong evidence in the radial profiles
for a break in slope indicative of 
an infall radius in the outer portion of the envelope 
probed by SCUBA.  This result directly contradicts the molecular line modeling
results towards B335 (Choi et al. 1995), which strongly
favors larger infall radii.  However, the molecular line
models do not take into account chemical effects and 
abundance gradients, which can affect the shape 
of the blue asymmetry profile (Rawlings \& Yates 2001).  Unfortunately,
the best fit power laws cannot be directly used in Monte Carlo molecular
line modeling without an assumption about the velocity field along the
line of sight.

\subsection{Mass Determinations}

The mass in the envelope within a 120\as\ aperture was calculated using
\begin{equation}
M_{env}^{120} = \int_{r_i}^{r(60\as )} \mu m_H n(r) 4 \pi r^2 dr .
\end{equation}
The average mass is $2.6 \pm 1.3$ \msun\ for the seven Class 0 best fit power laws,
$3.5$ times higher than the average \ppc\ mass
($0.8 \pm 0.1$ \msun ) within the same aperture (cf. Paper II).
The average mass within a fixed outer radius of $30,000$ AU is 
$9.5 \pm 5.1 $ \msun\ for Class 0 sources compared to $2.7 \pm 0.3$ \msun\
for \ppcs .  
If these Class 0 sources are the next stage of evolution,
they clearly have most of their mass in the envelope, as originally
proposed by Andr\`e et al. (1993).

In the absence of a realistic model for \Tdr,
masses are usually determined from submillimeter observations using
an isothermal approximation. 
To facilitate mass determinations for sources without models, we
explored the most suitable temperature to use.
We calculated the isothermal dust temperature that yields the same mass as
the detailed model based on the measured 850 \micron\ flux in a 120\as\ beam, 
$S_{850}^{120}$, using
\begin{equation}
T_{iso} = \frac{h\nu }{k} \; \left( 1 + \ln{\frac{2h\nu ^3 M_{env} \kappa _{\nu}}
                                                       {S_{850}^{120} c^2 D^2} }
                                   \right)^{-1} .
\end{equation}
The results are listed in Table 4. Most are 11-13 K, but the most
luminous source, L483, has $T_{iso} = 18$ K.
L1448C is excluded due to confusion from the northern sources 
in the 120\as\ aperture.  
The mean value is $13.8 \pm 2.4$ K.
The sources with $p \sim 1$ (L1527, L483) have higher $T_{iso}$
($15.0$K and $18.0$K respectively).  
A similar calculation can be made for the \ppcs\ modeled in Paper II
using the mass of a Bonnor-Ebert sphere within the 120\as\ aperture.
For the three \ppcs , the average isothermal temperature is
$11.1 \pm 1.2$ ($12.4$ K L1544, $10.9$ K L1689B, $10.0$K L1512),
slightly lower than for Class 0 sources with an internal luminosity source.

The model dust mass can be compared to the virial mass
calculated using optically thin linewidths (e.g., \nthp\ and
\hcopi; see Table 8 in Paper I).  The virial mass estimates in
Paper I assumed a constant density distribution in the envelope; however,
the virial mass can be corrected for the best fit power law density distribution
(cf. Bertoldi \& McKee 1992) using 
\begin{equation}
M_{virial}^{\theta _{ap}} = \frac{  (5-2p) D \theta _{ap} \mean{v^2}}
				 {2  (3 - p) G} , \;\; p < 2.5 \; ,
\end{equation}
where $\theta _{ap}$ is the FWHM size of the aperture in which the dust mass 
was determined, $D$ is the distance, and \mean{v^2} is the 3-D velocity
dispersion:

\begin{equation}
\mean{v^2} = 3 \left[\frac{kT_{iso}}{\mu m_H} + \left(\frac{\Delta v^2}{8 ln2} - 
\frac{kT_{iso}}{m_{amu}m_H}\right) \right],
\end{equation}
where $\Delta v$ is the FWHM linewidth of an optically thin line, 
$\mu$ is the mean molecular mass, and $m_{amu}$ is the molecular
mass of the species whose linewidth is used.

The mean of the ratio of the virial mass to the model dust mass in a 120\as\ aperture 
is $2.1 \pm 0.6$. Uncertainties of up to a factor of $10$ in the opacity exist between
different dust models (Ossenkopf \& Henning 1994).  
We regard this agreement as
encouraging evidence that OH5 opacities describe the dust in the cores well
and that the virial theorem, properly applied, gives good mass estimates.
This ratio is also consistent with
the ratio of virial mass to model dust mass for a sample of deeply embedded high mass
star forming cores associated with water masers ($2.4 \pm 1.4$, Evans et al. 2002).
A factor of $2$ decrease in $\kappa _\nu(850)$ from the OH5 value would bring the
dust model mass and virial mass into agreement on average.

\subsection{Spectral Indices}

The spectral index, 
$\alpha_{450/850}^{120}$, was calculated 
from the model fluxes at 450 and 850 \micron\ in a 120\as\ aperture 
(L1448C excluded due to confusion) using
\begin{equation}
\alpha_{450/850}^{120} = \frac{ \log ( S_{450}^{mod} /  S_{850}^{mod} ) }
                              { \log (850 / 450 ) } .
\end{equation}
The model spectral index calculated within a 120\as\
aperture agrees extremely well with the observed spectral index from Paper I; 
$\left< \alpha_{450/850}^{120}(model) / \alpha_{450/850}^{120}(observed) \right> 
= 0.96 \pm 0.18$.  
The model spectral indices are within the observed error bars;
however, the total uncertainty on the 450 \micron\ flux ($\sim 50\% $) makes the
observed spectral index fairly uncertain.
OH5 opacities are also successful in reproducing the observed submillimeter
spectral indices for Class 0 sources.

We also considered possible changes in the spectral index as a function of
impact parameter.  To avoid calibration uncertainties 
we compare the {\it normalized} specific intensity, 
$I_{450}^{norm}(b)/I_{850}^{norm}(b)$, 
for the best fit models to the observed ratios (Figure 13).
The variations in the model ratio match those in the observed ratio.  Apparent
variations in the spectral index can be removed  by taking into account 
beam effects and a realistic temperature distribution.  For
example, the first sidelobe at 450\micron\ results in a large increase in the 
ratio.  The lower signal-to-noise in outer
annuli would mask any subtle variations in the dust opacity.  L483 is a clear 
exception as the model specific intensity ratio does not decrease as fast as 
the observed ratio.  The 450 \micron\ model is too flat at large radii.  
However, this discrepancy may be caused by our use of a spherical model 
on an aspherical source rather than an actual variation in the dust 
properties. 
There is thus little evidence for variations of the dust opacity with 
radius.  This result contrasts with some earlier work (e.g., Visser et al. 1998,
Johnstone \& Bally 1999) in regions forming more massive stars, but
agrees with the conclusions of Chandler \& Richer (2000), who were
studying regions similar to those in this study.
It is extremely important to use a realistic beam profile and \Tdr\
for modeling both the 850 and 450 \micron\ SCUBA maps.  

\subsection{The ISRF and \lint }

The contribution from the ISRF can be important to the heating of
the region of the outer envelope probed by SCUBA observations of low luminosity
cores.  Scaling the \uv\ to \fir\ portion of the ISRF by a factor of 3 in 
either direction resulted in changes of the best fit power by 
factors of $\Delta p = \pm 0.2$.  Every source profile except L1448C 
was better fitted by a lower strength of the ISRF, $\sisrf = 0.3$, in agreement
with models of \ppc s (Paper II).  Modest shielding of the ISRF by surrounding
gas in the molecular cloud surrounding the dense core could account for this.  
L1448C is forming in a much more
crowded region and may be subjected to a stronger ISRF.

Our models provide an estimate of the internal luminosity of the central
protostar.  For all of our sources, the internal luminosity accounts for nearly
all of the observed luminosity ($\left< L_{int}/L_{bol}^{obs}
\right> = 0.95 \pm 0.12$).
While the heating from the ISRF is important for the temperature structure
of the outer envelope, it does not contribute significantly to the overall
observed luminosity of the Class 0 sources we modeled.  There is a wide
range of internal luminosities modeled, from $1.0$ \lsun\ to $13.0$ \lsun, 
with an average internal luminosity of $4.0$ \lsun . 
For each of the best fit models, the flux at 850 \micron\ and the
observed bolometric luminosity were fitted simultaneously.
For sources with SEDs not observed shortward of 60 \micron , the model
internal luminosity is a lower limit.  All of the luminosities were
calculated using the distances from Paper I.  Since the distance to
many globules are uncertain to 50\%, a true determination of
the internal luminosity is uncertain to 100\%; however,
the model and observed internal luminosities are consistent for the
distance adopted.

Interestingly, the most luminous source, L483, has about the same 
envelope mass as the others. The greater luminosity should reflect either
a higher stellar mass or a higher accretion rate. If the former, it would
imply a higher starting mass for the condensation; if the latter, it might
be reflected in the linewidths. In fact, the linewidth of the \nthp\ line
is similar to that of the other sources. Alternatively, the accretion
might be in a transient high state, similar to an FU Orionis event.

\subsection{Caveats and Future Work}

While our models take into account heating from a central source,
heating from the ISRF, realistic beam effects, and simulated chopping, they
cannot effectively model asymmetries seen in the dust continuum maps, 
flattening due to rotation or effects of magnetic fields, and clearing 
of material in outflow cavities.  
Five of the cores modeled (B335, B228, L723, IRAS03282+3035, L1448C) appear 
sufficiently symmetric that the most important effect not included would
be lower densities in outflow cavities.  
Harvey et al. (2001) found a strong asymmetry in the
(H -- K) colors along the outflow directions in B335, indicative of clearing of
material along the outflow axes. The resulting best fit models included an outflow
opening angle of 35\degree\ to 45\degree . While no such asymmetry is observed in 
the submillimeter continuum maps of B335 (Paper I), extensions along the 
outflow directions (e.g., L1527, L723) and perpendicular to the outflow
direction (e.g., L483) are observed in submillimeter maps of Class 0 sources.
The overall impact outflows have on the density structure of the envelope
will become clearer with finer resolution and higher sensitivity.
Future multi-dimensional modeling of the dust continuum emission 
should attempt to account for the effects of the outflow.    

We have neglected the emission from a disk.  Chandler \& Richer (2002) showed
that the flux from a disk is negligible compared to the
the total flux from the envelope of Class 0 sources; however, Class 0
sources may have a substantial fraction of emission from a compact component
within the central beam.  The interpretation of the density structure
of the outer envelope may be strongly affected when a centrally 
normalized radial profile is used.  
Only a few Class 0 sources have been observed at submillimeter wavelengths
with interferometers (cf. Brown et al. 2000) resulting in few constraints on 
submillimeter disk fluxes.  Observations of 2.7mm continuum towards
L1527 with OVRO and BIMA find a flux of $\sim 40$ mJy from a compact
component (Terebey et al. 1994, Shirley et al. unpublished observation).  
Using a model for an active disk ($T(r) \sim r^{-0.5}$; Butner et al. 1994), 
we find an upper limit of $0.7$ Jy at 850 \micron .  In this scenario, 
the disk accounts for up to $70\%$ of the flux within the central beam, 
decreasing $p$ by $\Delta p = -0.6$.  
It is likely that the 7\as\ aperture included flux from the envelope, thereby
overestimating the flux from a disk.  If the disk
has a steeper temperature power law ($T(r) \sim r^{-0.75}$), then
the flux contribution drops to $0.1$ Jy at 850 \micron\ and the change in 
the best fit power law model is $\Delta p \sim -0.1$.  As another example, the
Choi et al. (1996) Shu77 model fits the B335 radial profile when 
the disk flux equals the envelope flux within the central beam
($0.4$ Jy at 850 \micron ).  However, BIMA millimeter observations of B335 do
not support such a high disk flux (Shirley et al., unpublished observation).  
Constraints on the disk flux and modeling of BIMA observations of 
the dust continuum towards Class 0 protostars will be presented in a future paper.

\section{Conclusions}

We have modeled seven Class 0 sources using single power law and
Shu77 density distributions.  Power law models suitably fit the 850 \micron\ 
profiles and SED of all Class 0 sources.  Five sources with circular
contours are best fitted by a steep value of the power law index 
$p = 1.7 - 1.9$ 
(B335, B228, L723, IRAS03282+3035, L1448C), while two sources with aspherical
emission are fitted by flatter power law indices $p = 1.1- 1.2$ 
(L1527, L483).  
Uncertainties in the strength of the ISRF, \sisrf , and beam shape limit 
the accuracy in the power law index to $\pm 0.2$.  

The Shu77 model from Choi et al. (1995) does not fit 
the B335 radial profiles.  Smaller infall radii are able to
fit the profiles (B335, B228, L723, IRAS03282+3035, L1448C), 
but the infall radius is within the central 450 \micron\ 
beam, effectively making the density distribution appear like
a single power law throughout the region of the envelope probed
by SCUBA.

	The average mass within a 120\as\ aperture is $2.6$\msun\ and
is reasonably consistent with virial mass estimates and observed mass estimates
from Paper I and models of the initial conditions from Paper I.  
We find little evidence for variations in the dust opacity with radius.  
OH5 dust reproduces the observed spectral index on average and provides
a good fit to many SEDs. In addition, it leads to masses determined from
dust emission that are consistent with virial masses to a factor of 2.
Heating from the ISRF is very important for correctly
interpreting the temperature profile of the outer envelope of low
mass star forming cores but does not significantly contribute to the
total bolometric luminosity.  The dust models constrain the internal
luminosity of Class 0 protostars, but distances to isolated
cores remain the largest uncertainty in determining accurate
masses and luminosities.

	Outflow cavities and asymmetrical density
distributions should be modeled using higher dimensional dust modeling.
In particular, L483 and L1527 have large asymmetries in the dust continuum
emission that cannot be modeled with a one dimensional code.

	The presence of a disk within the central beam may affect the
interpretation of the best fit density distribution in the outer 
envelope, decreasing $p$ by as much as $\Delta p = -0.6$.

	We can now use a more realistic $n(r)$ and $T_d(r)$ in
Monte Carlo molecular line radiative transfer models to test
infall models (which provide $v(r)$) and the origin of line asymmetries, 
to test predictions
of chemical models, to investigate amounts of depletion, and to 
improve estimates of the ionization fraction.  

Our primary conclusion is that a simple power law for the density distribution
fits all of the Class 0 sources that we have considered, while Shu77
models with substantial $r_{inf}$ do not fit.  Firmer conclusions await 
stronger constraints on the submillimeter flux from a possible disk
and modeling of interferometric observations of Class 0 protostars.

\section*{Acknowledgments}

We are grateful to L. Mundy for providing the computer code used
for beam convolution and solution of Equation (1) and for stimulating
discussions about the effects of disk.  We thank
Steve Doty for useful discussions and consistency checks of our 1D models.  
We thank Chad Young for his help simulating effects of a disk.
We are grateful to the referee, Antonella Natta, who made many helpful 
suggestions. We thank the State of Texas and NASA (Grants NAG5-7203 and 
NAG5-10488) for support. NJE thanks the Fulbright Program and PPARC for 
support while at University College London and NWO and NOVA for support in Leiden.
The JCMT is operated by the Joint Astronomy Centre on behalf of the Particle
Physics and Astronomy Research Council of the United Kingdom, The Netherlands 
Organization for Scientific Research and the National Research Council of 
Canada.   



\begin{figure}
\figurenum{1}
\plotone{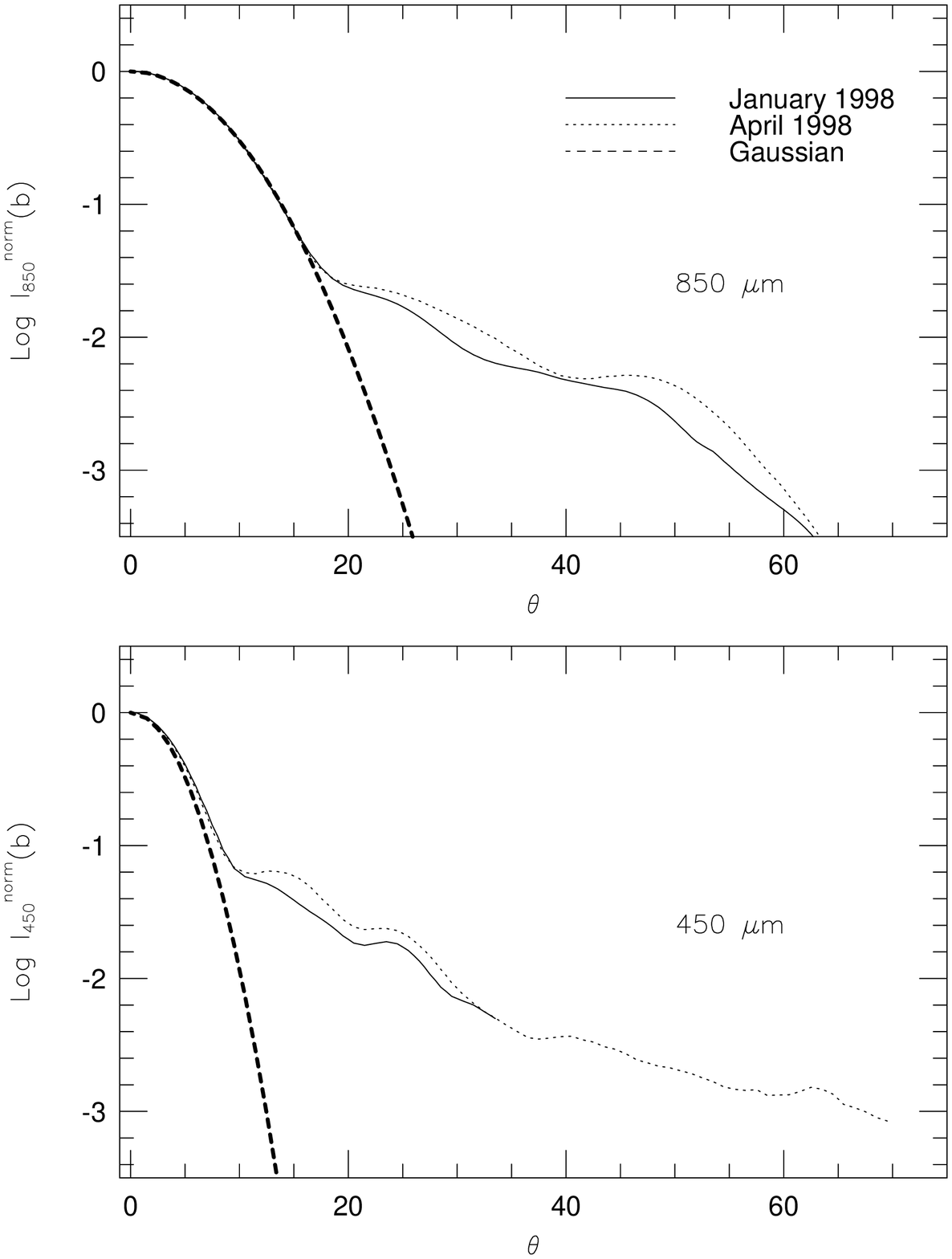}
\figcaption{
Beam profiles used for modeling Class 0 sources.  The beam profile from January 1998,
April 1998, and gaussian beams with the FWHMs reported in Paper I are shown.
Since CRL618 is a weak calibrator, the January 1998 beam was extrapolated beyond
34\as\ using the April 1998 beam.}
\end{figure}

\begin{figure}
\figurenum{2}
\plotone{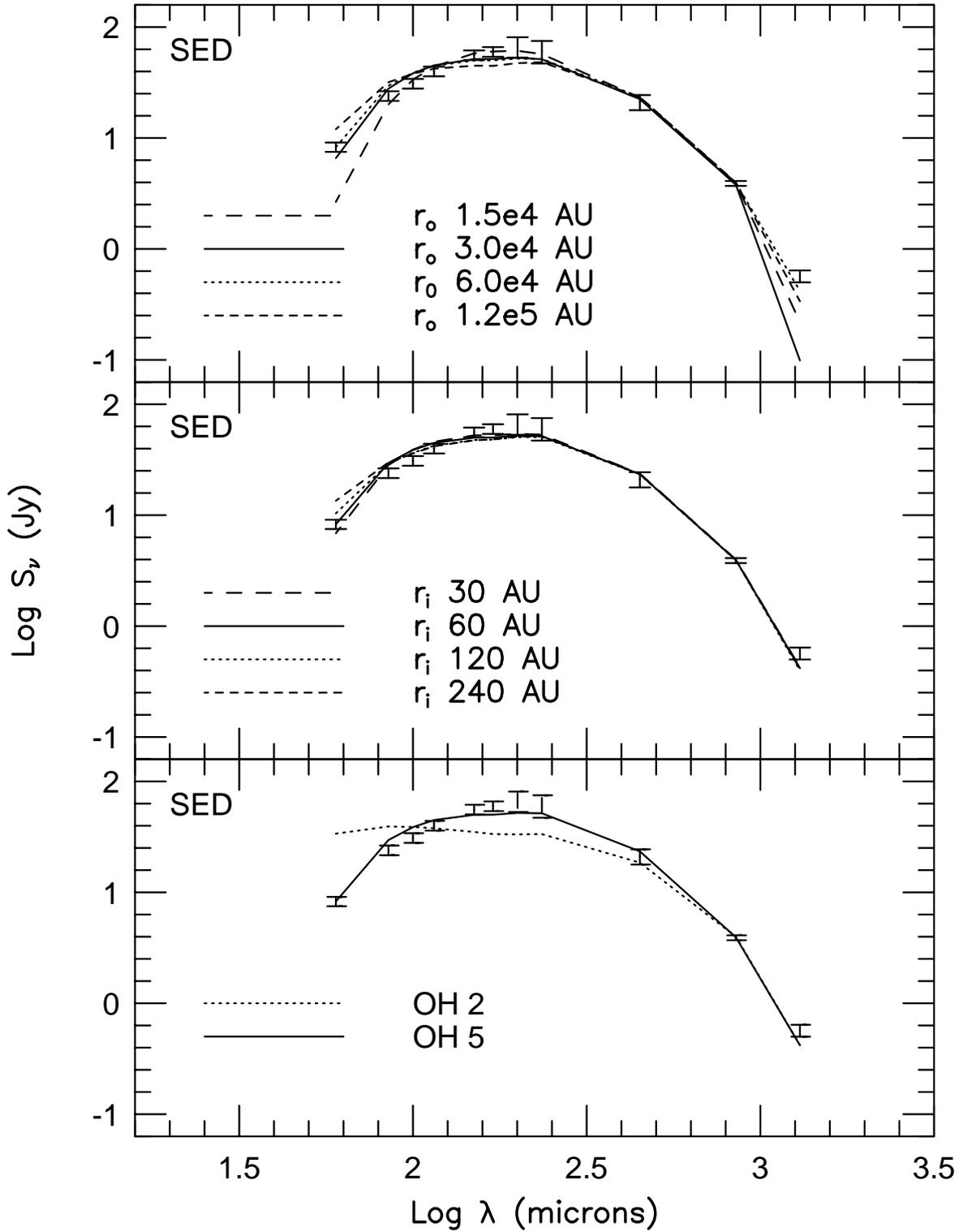}
\figcaption{
B335 Model SEDs for power laws with p=2.0 for various $r_o$, $r_i$, and 
$\kappa _{\nu}$.  Only the model SEDs are plotted since effects to the
radial profiles are shown to be negligible.
}
\end{figure}

\begin{figure}
\figurenum{3}
\plotone{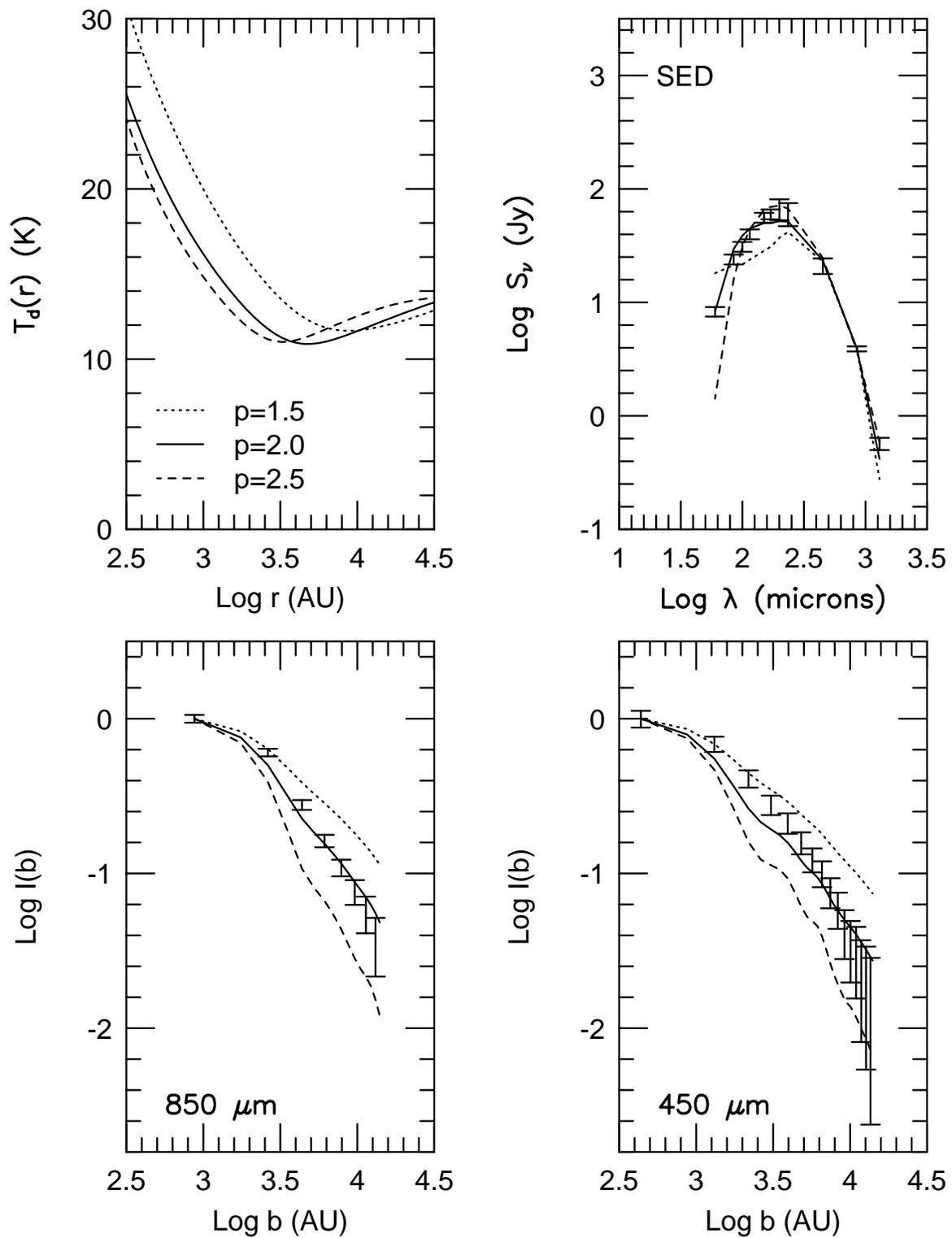}
\figcaption{
B335 power Law models with $p = 1.5$, 2.0, and 2.5.  The
$p = 2.0$ model is the best fit to the SED and radial profiles.  The
model parameters were: OH5 dust opacities, \sisrf\ = $1.0$, 
$r_i = 60$ AU, and $r_o = 60,000$AU for all models.  
}
\end{figure}

\begin{figure}
\figurenum{4}
\plotone{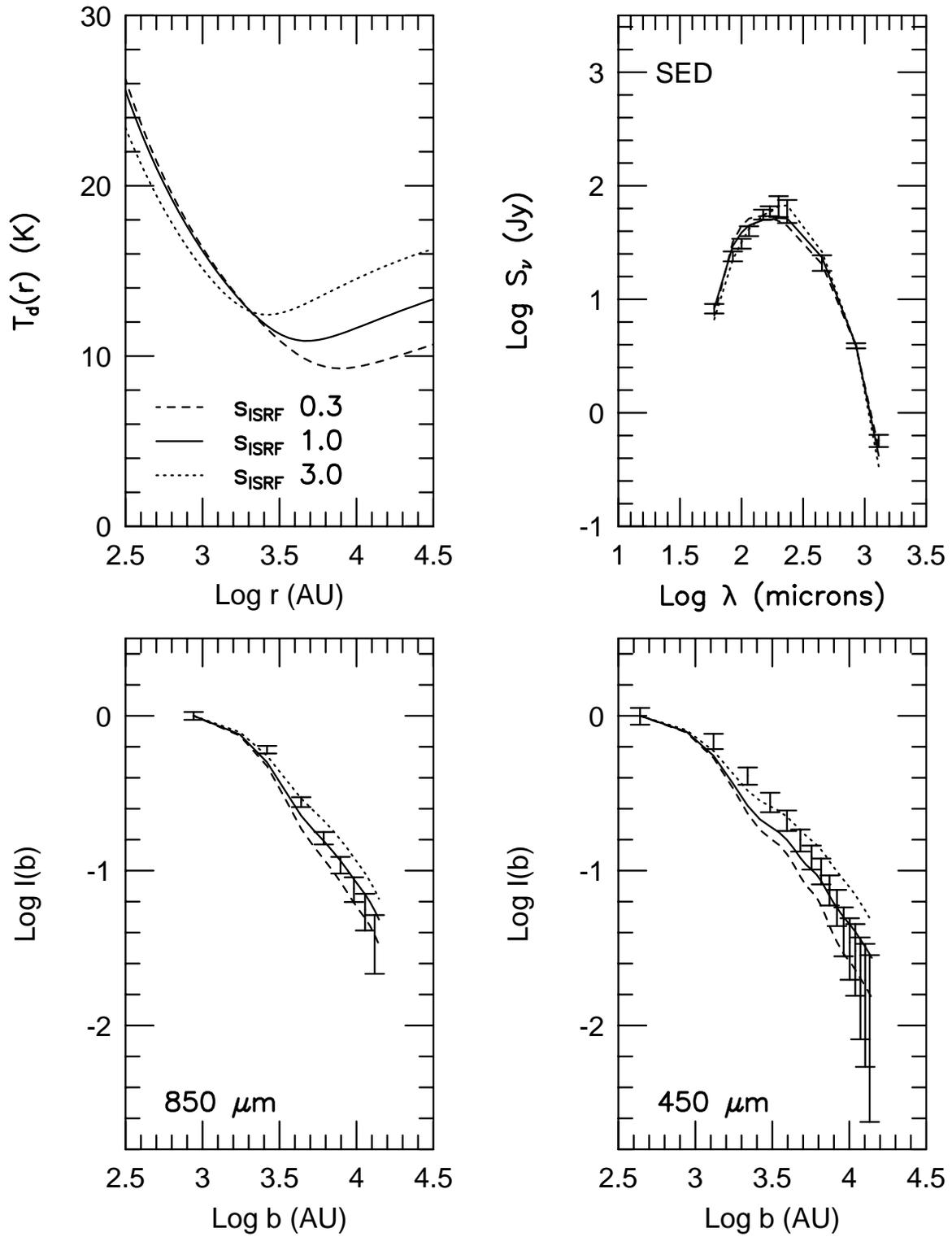}
\figcaption{
Effects of changing the strength of the ISRF.  Three $p = 2$ power law models of B335 
are shown with \sisrf\ $ = 0.3, 1.0,$ and $3.0$.  
}
\end{figure}

\begin{figure}
\figurenum{5}
\plotone{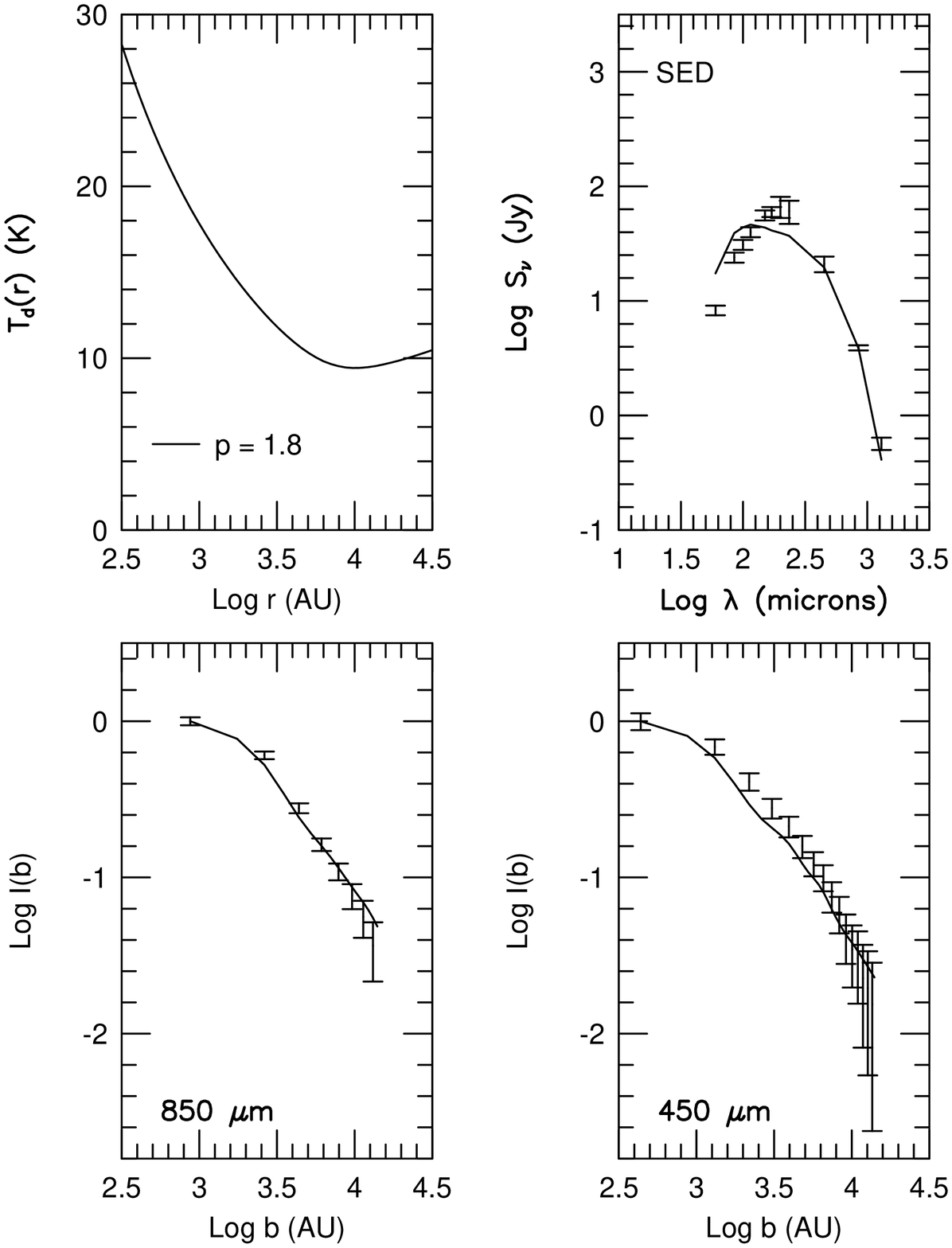}
\figcaption{
The best fit power law model for B335.  The parameters of the best fit
were: $p = 1.8$, $n_f = 1.5 \times 10^6$ \cmv , $L_{int} = 3.3$\lsun , and
\sisrf\ $= 0.3$.
}
\end{figure}

\begin{figure}
\figurenum{6}
\plotone{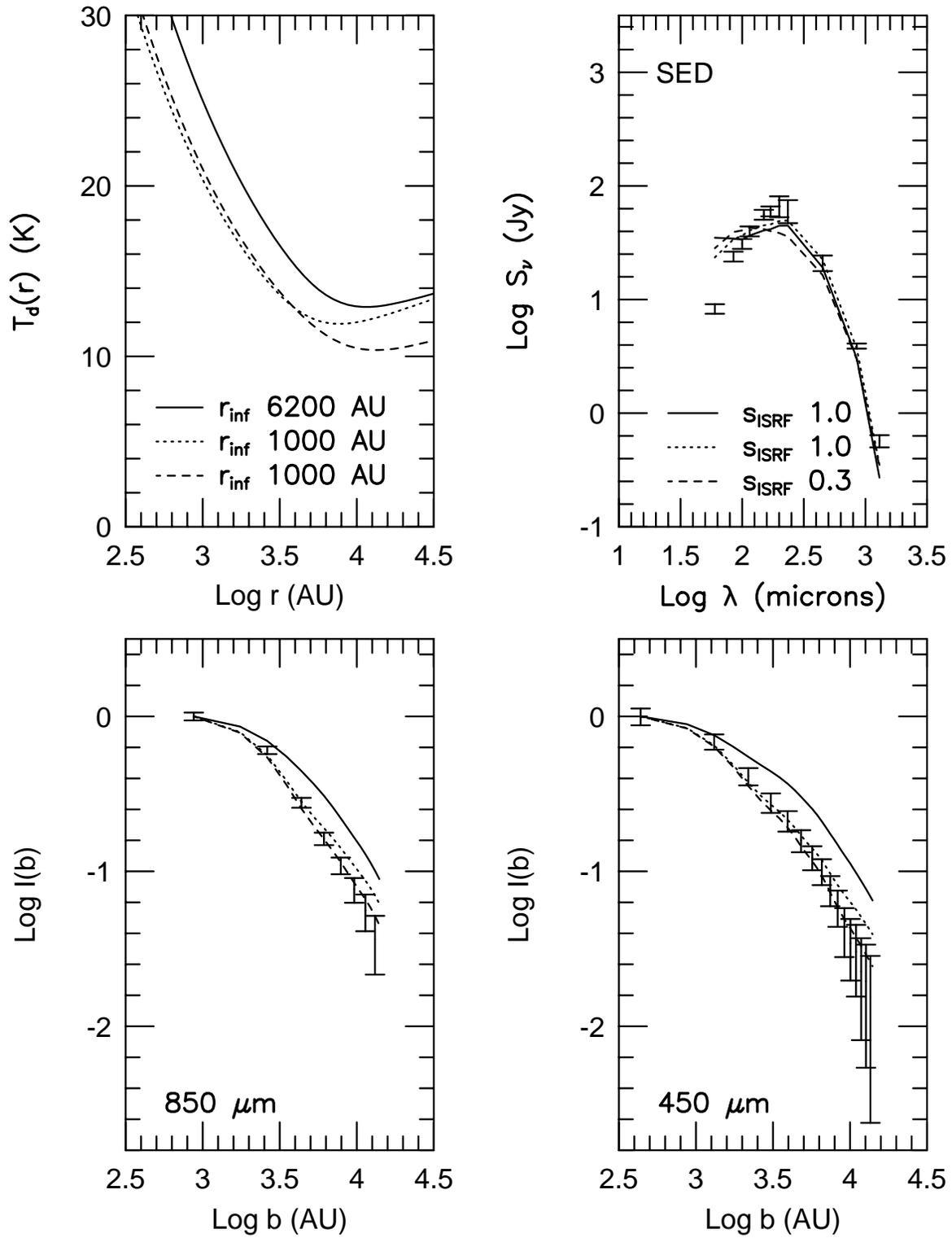}
\figcaption{
Shu77 models for B335.  Three models are shown: $r_{inf} = 6200$ AU and 
\sisrf\ $= 1.0$, $r_{inf} = 1000$ AU and \sisrf\ $= 1.0$, 
and $r_{inf} = 1000$ AU and \sisrf\ $= 0.3$.  The smaller infall radius of 1000 AU
is a better fit than the best fit from Choi et al. (1995).
}
\end{figure}

\begin{figure}
\figurenum{7}
\plotone{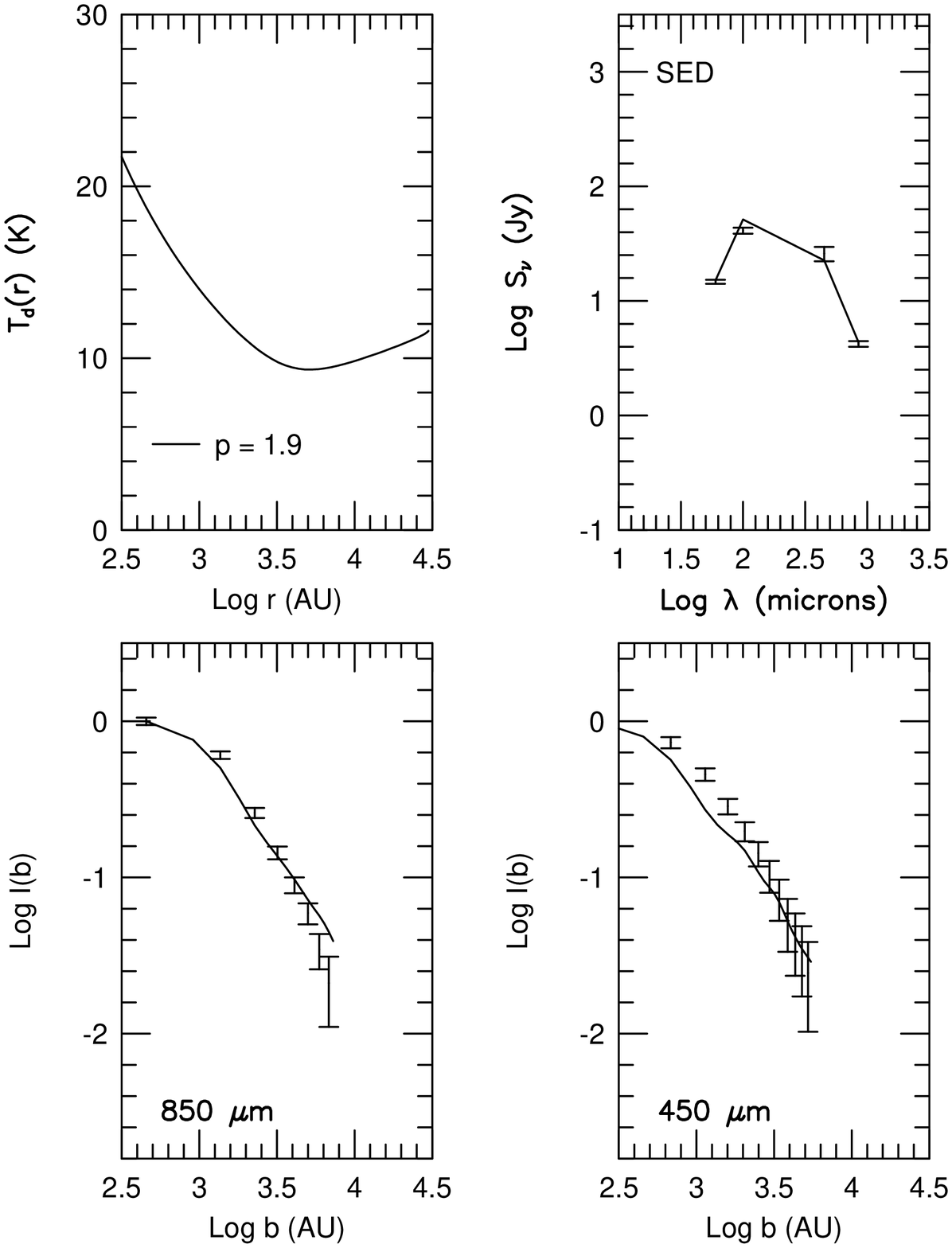}
\figcaption{
The best fit power law model for B228.  The parameters of the best fit
were: $p = 1.9$, $n_f = 1.2 \times 10^6$ \cmv , $L_{int} = 1.0$\lsun , and
\sisrf\ $= 0.3$.
}
\end{figure}

\begin{figure}
\figurenum{8}
\plotone{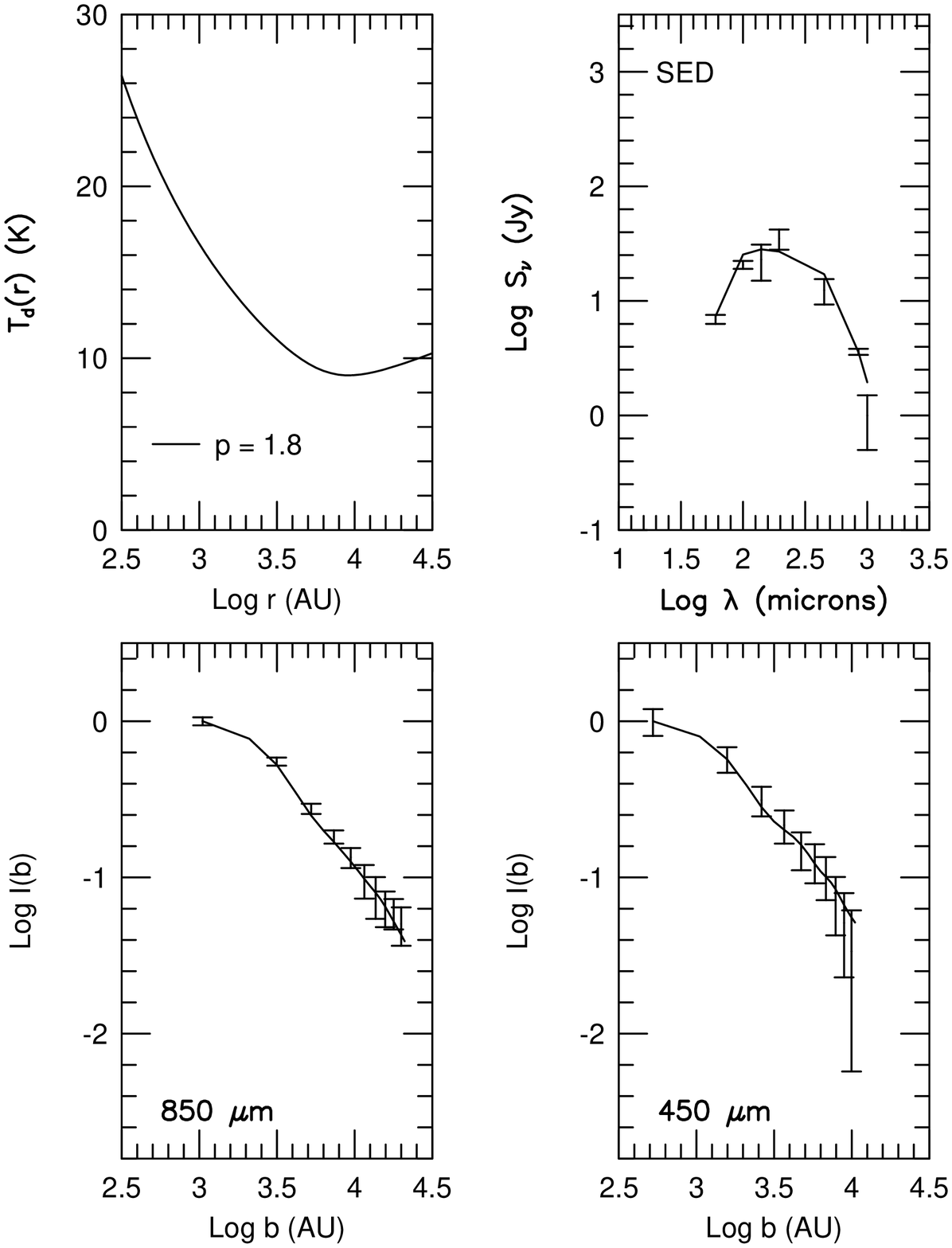}
\figcaption{
The best fit power law model for L723.  The parameters of the best fit
were: $p = 1.8$, $n_f = 1.8 \times 10^6$ \cmv , $L_{int} = 2.6$\lsun , and
\sisrf\ $= 0.3$.
}
\end{figure}

\begin{figure}
\figurenum{9}
\plotone{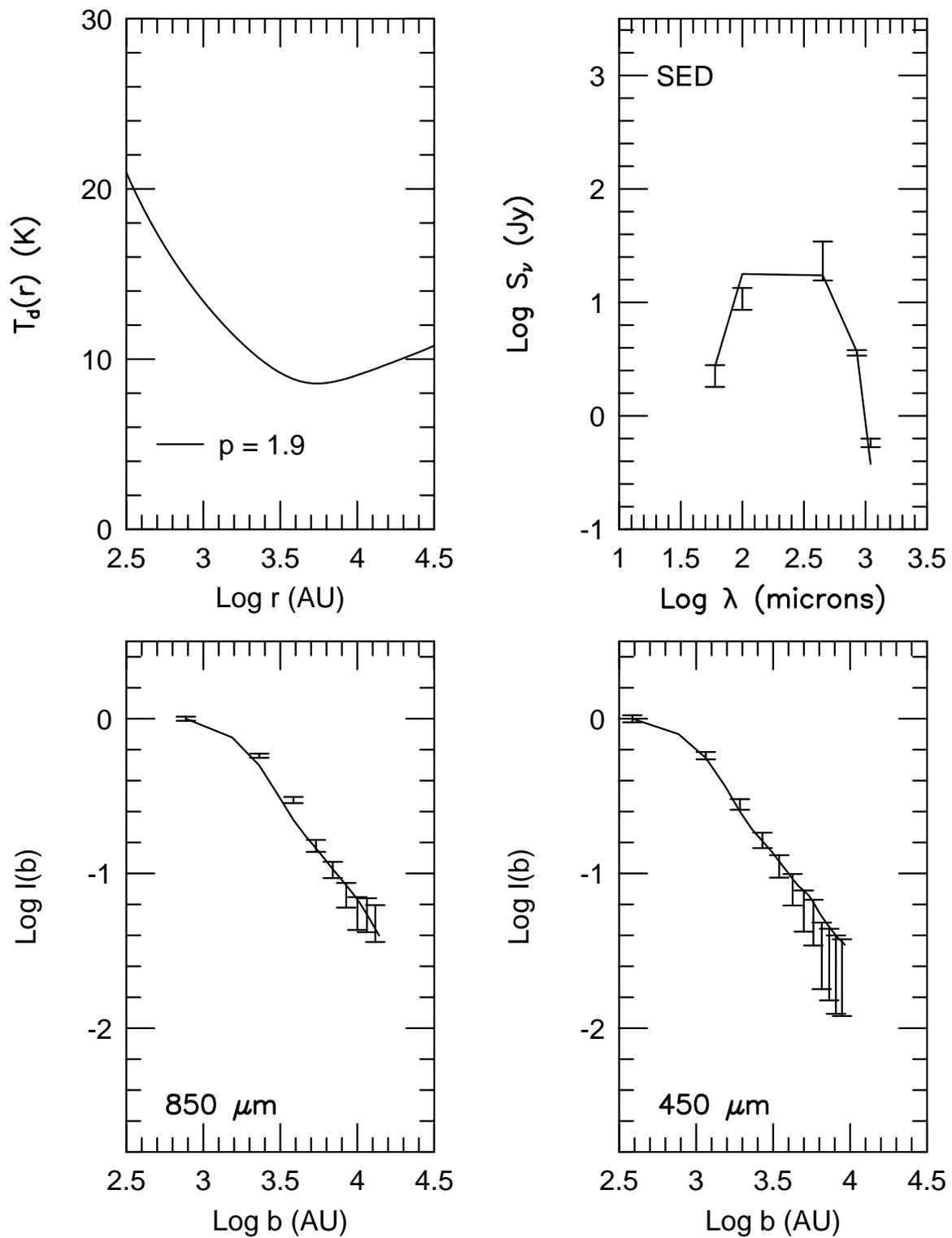}
\figcaption{
The best fit power law model for IRAS03282+3035.  The parameters of the best fit
were: $p = 1.9$, $n_f = 1.9 \times 10^6$ \cmv , $L_{int} = 1.0$\lsun , and
\sisrf\ $= 0.3$.
}
\end{figure}

\begin{figure}
\figurenum{10}
\plotone{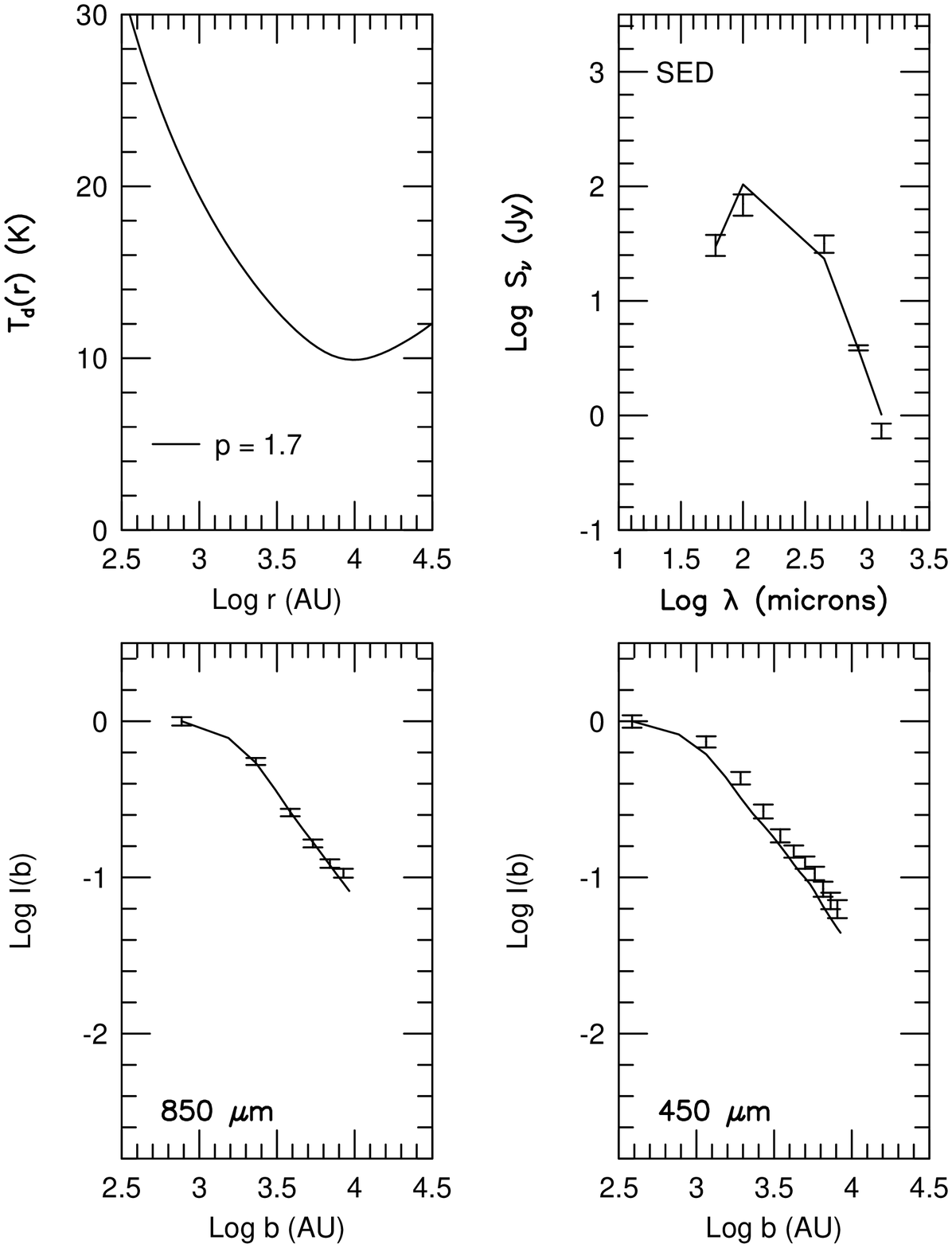}
\figcaption{
The best fit power law model for L1448C.  The parameters of the best fit
were: $p = 1.7$, $n_f = 2.6 \times 10^6$ \cmv , $L_{int} = 5.9$\lsun , and
\sisrf\ $= 1.0$.
}
\end{figure}

\begin{figure}
\figurenum{11}
\plotone{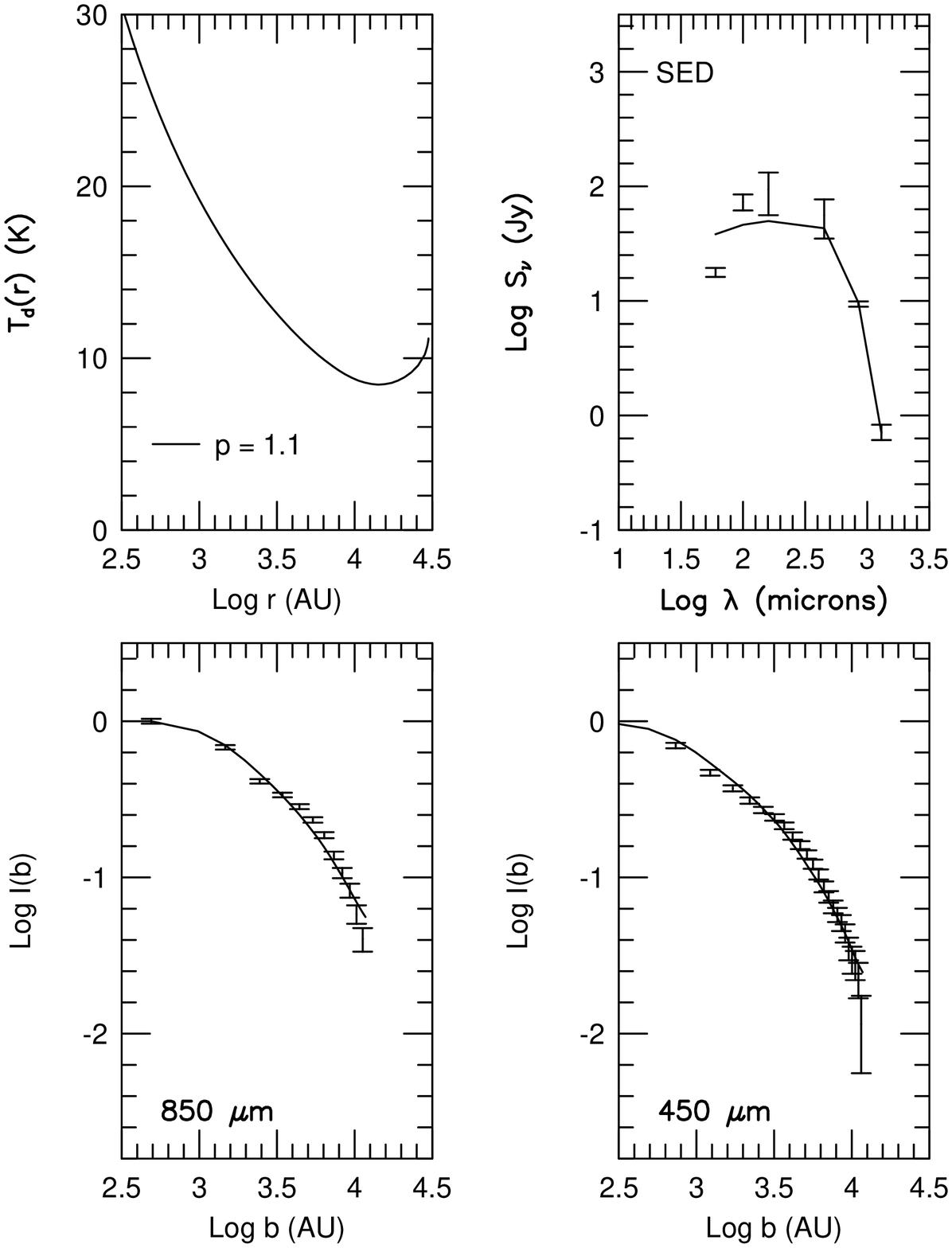}
\figcaption{
The best fit power law model for L1527.  The parameters of the best fit
were: $p = 1.1$, $n_f = 6.5 \times 10^5$ \cmv , $L_{int} = 2.2$\lsun , and
\sisrf\ $= 0.3$.
}
\end{figure}

\begin{figure}
\figurenum{12}
\plotone{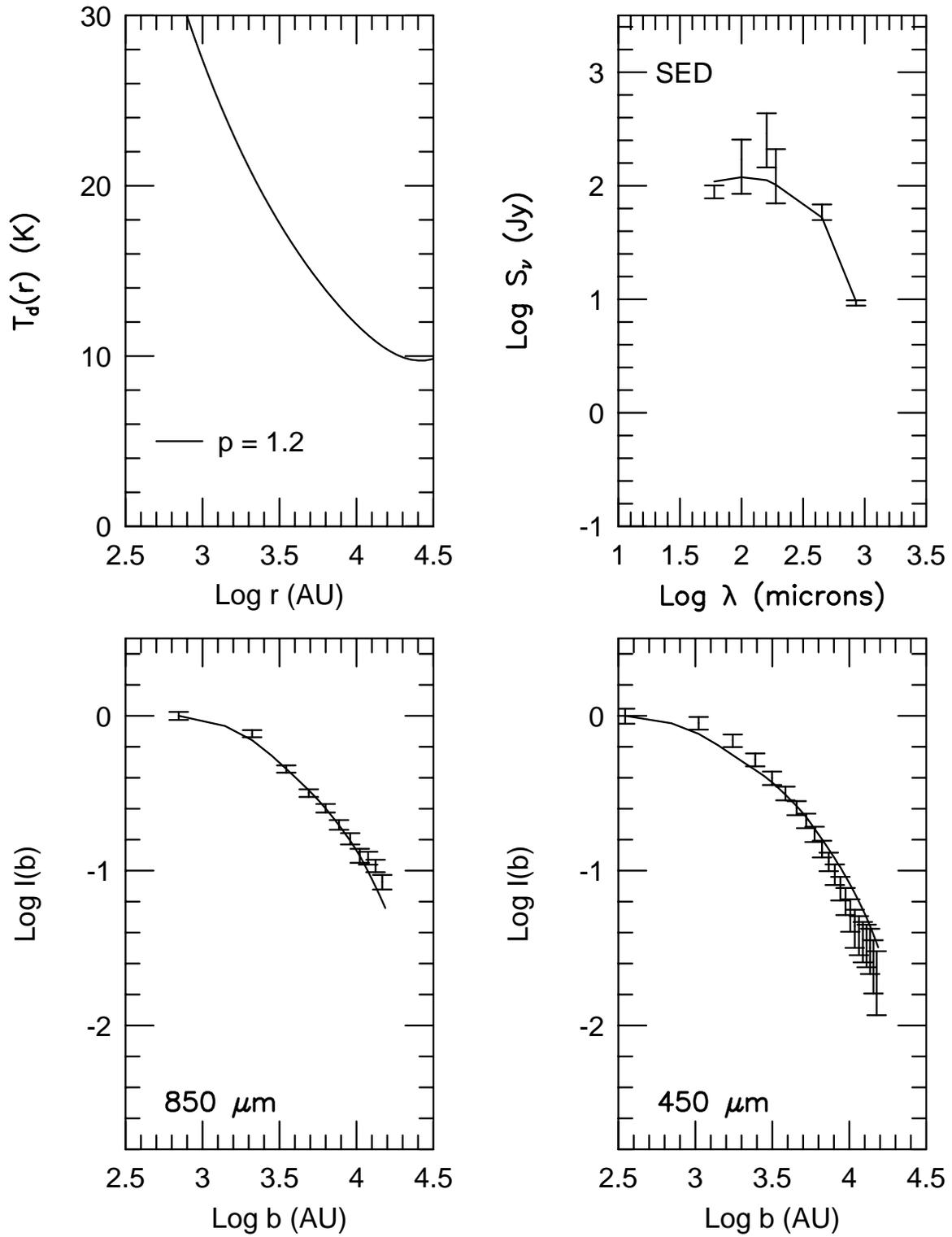}
\figcaption{
The best fit power law model for L483.  The parameters of the best fit
were: $p = 1.2$, $n_f = 6.0 \times 10^5$ \cmv , $L_{int} = 13.0$\lsun , and
\sisrf\ $= 0.3$.
}
\end{figure}

\begin{figure}
\figurenum{13}
\plotone{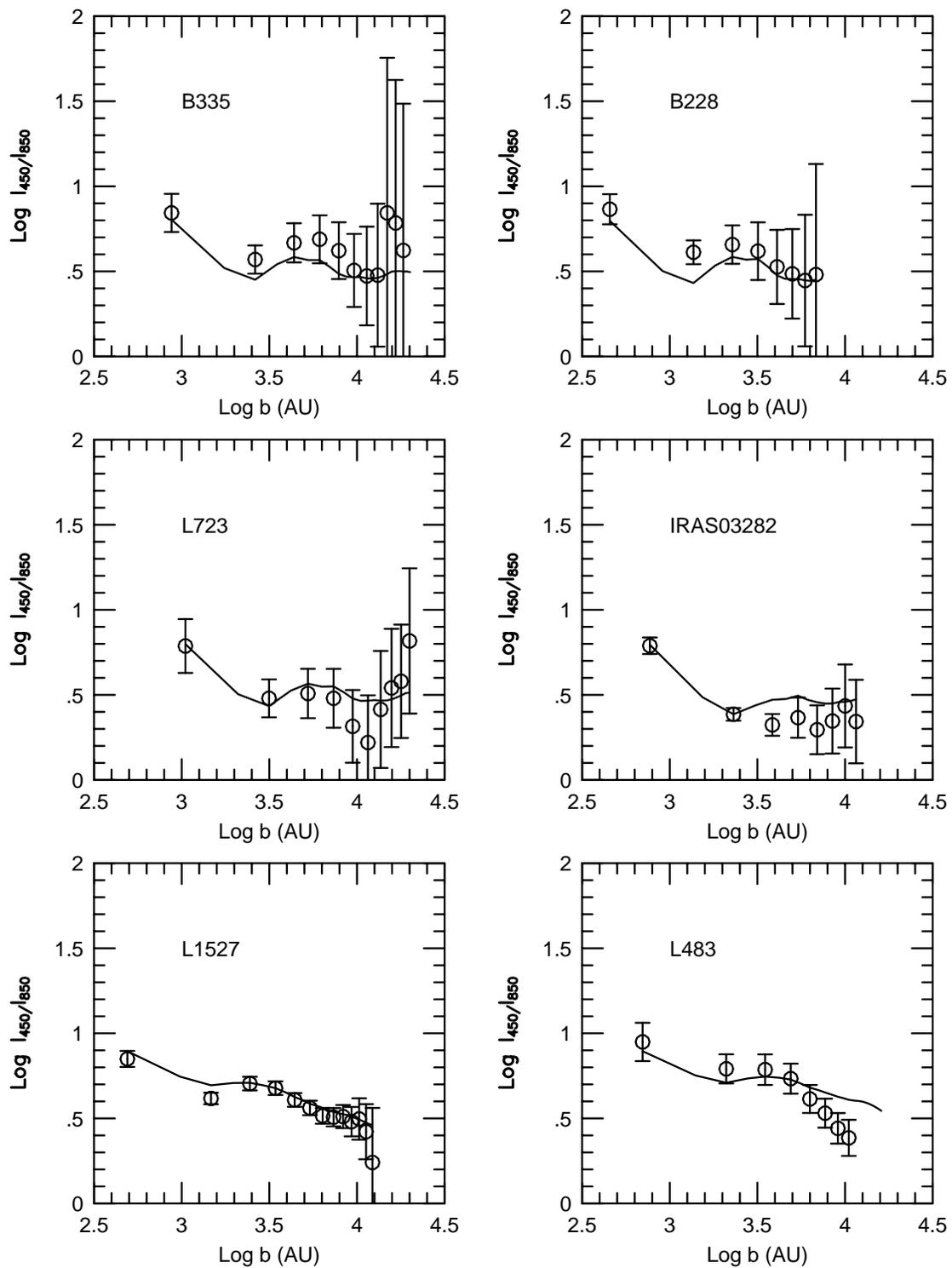}
\figcaption{
The ratio of specific intensity, $I_{450}^{norm}/I_{850}^{norm}$  
for the best fit power law models of isolated Class 0 sources (solid line).  
Points are spaced at the half beam resolution at 850 \micron .  
}
\end{figure}


\begin{center} 
\begin{table}
\caption{Power Law Models \label{physmods}}
\begin{tabular}{llllllllllll}
\tableline\tableline
Source  & $p$    & $n_f$              & $r_i$ & $r_o$   & $\kappa _\nu$ &
$L_{int}$ & \sisrf\        & $\chi ^2_{850}$ & $\chi ^2_{450}$ & $\chi ^2_{SED}$\tablenotemark{a} 
& Notes \\
&        & ($\cmv$)           & (AU)  & (AU)    &               &
(\lsun )&                  &                     &                  &              &       \\
\tableline
B335    & \\
& 2.0     & $1.7 \times 10^6 $ & 60    & 60,000  &  OH5 &  2.5 & 1.0 & 5.9  & 3.9  & 2.2 & Test Model\\
& 2.0	  & $1.7 \times 10^6 $ & 30    & 60,000  &  OH5 &  2.5 & 1.0 & 6.8  & 5.0  & 2.1 \\
& 2.0     & $1.7 \times 10^6 $ & 120   & 60,000  &  OH5 &  2.5 & 1.0 & 5.4  & 3.4  & 2.7 \\ 
& 2.0     & $1.2 \times 10^6 $ & 240   & 60,000  &  OH5 &  2.7 & 1.0 & 4.4  & 2.1  & 5.9 \\ 
& 2.0     & $1.2 \times 10^6 $ & 60    & 120,000 &  OH5 &  2.5 & 1.0 & 5.0  & 3.5  & 5.7 \\ 
& 2.0     & $2.0 \times 10^6 $ & 60    & 30,000  &  OH5 &  2.5 & 1.0 & 6.5  & 3.8  & 5.8 \\ 
& 2.0     & $3.0 \times 10^6 $ & 60    & 15,000  &  OH5 &  2.5 & 1.0 & 16.9 & 10.8 & 7.3 \\
& 1.8     & $2.5 \times 10^6 $ & 60    & 15,000  &  OH5 &  2.5 & 1.0 & 5.4  & 3.9  & 5.8 \\
& 1.5     & $5.0 \times 10^5 $ & 60    & 60,000  &  OH5 &  2.8 & 1.0 & 79.4 & 34.5 & 22.6 \\
& 2.5     & $2.8 \times 10^6 $ & 60    & 60,000  &  OH5 &  2.5 & 1.0 & 47.1 & 19.7 & 8.5 \\ 
& 2.0     & $1.2 \times 10^6 $ & 60    & 60,000  &  OH5 &  1.5 & 3.0 & 14.7 & 9.5  & 1.7 \\ 
& 2.2     & $1.7 \times 10^6 $ & 60    & 60,000  &  OH5 &  1.5 & 3.0 & 5.6  & 3.4  & 5.6 \\ 
& 2.0     & $2.3 \times 10^6 $ & 60    & 60,000  &  OH5 &  3.0 & 0.3 & 12.2 & 8.9  & 4.2 \\ 
& 1.8     & $1.5 \times 10^6 $ & 60    & 60,000  &  OH5 &  3.3 & 0.3 & 3.8  & 2.9  & 19.6 & Best $\chi^2_r$ \\ 
& 2.0     & $6.0 \times 10^5 $ & 60    & 60,000  &  OH2 &  4.4 & 1.0 & 6.2  & 3.8  & 101  \\ 
B228      & \\ 
& 2.1     & $1.2 \times 10^6 $ & 60    & 30,000  &  OH5 &  1.0 & 1.0 & 11.4 & 8.6  & 30.7 \\ 
& 1.9     & $1.2 \times 10^6 $ & 60    & 30,000  &  OH5 &  1.0 & 0.3 & 8.8  & 6.4  & 4.8  & Best $\chi^2_{tot}$   \\
L723      & \\
& 2.0     & $2.0 \times 10^6$  & 60    & 60,000  &  OH5 &  2.0 & 1.0 & 1.1  & 1.3  & 9.6  \\ 
& 1.8     & $1.8 \times 10^6$  & 60    & 60,000  &  OH5 &  2.6 & 0.3 & 0.8  & 0.5  & 3.4  & Best $ \chi^2_{tot}$ \\
& 1.8     & $2.4 \times 10^6$  & 60    & 30,000  &  OH5 &  2.6 & 0.3 & 0.8  & 0.6  & 3.7  \\ 
IRAS03282\tablenotemark{b} & \\ 
& 2.1     & $1.9 \times 10^6$  & 60    & 45,000  &  OH5 &  1.0 & 1.0 & 13.5 & 3.5  & 4.6  \\ 
& 1.9	  & $1.9 \times 10^6$  & 60    & 45,000  &  OH5 &  1.0 & 0.3 & 10.2 & 1.2  & 5.2  & Best $\chi^2_r$ \\
L1448C    & \\
& 1.7	  & $2.6 \times 10^6$  & 60    & 45,000  &  OH5 & 5.9  & 1.0 & 4.7  & 10.4 & 2.0  & Best $ \chi^2_{tot}$ \\ 
& 1.6	  & $2.8  \times 10^6$ & 60    & 45,000  &  OH5 & 5.9  & 0.3 & 8.3  & 9.4  & 9.3  & \\
L1527     & \\
& 1.5	  & $1.15 \times 10^5$ & 60    & 30,000  &  OH5 &  1.8 & 1.0 & 76.7 & 39.6 & 13.7 \\ 
& 1.0     & $3.5 \times 10^5$  & 60    & 30,000  &  OH5 &  2.1 & 1.0 & 52.3 & 60.3 & 25.6 \\ 
& 1.1     & $6.5 \times 10^5$  & 60    & 30,000  &  OH5 &  2.2 & 0.3 & 9.2  & 5.4  & 28.3 & Best $\chi^2_r$ \\
L483      & \\
& 1.3     & $6.0 \times 10^5$  & 60    & 45,000  &  OH5 & 13.0 & 1.0 & 8.9  & 5.1  & 1.2  \\
& 1.2     & $6.0 \times 10^5$  & 60    & 45,000  &  OH5 & 13.0 & 0.3 & 6.1  & 5.2  & 1.0  & Best $\chi^2_{tot}$ \\
\tableline
\end{tabular}
\tablenotetext{a}{$\chi ^2_{SED}$ calculated using all flux points listed in Tables 5, 6, and 7
of Shirley et al. 2000 with $\lambda \geq 60 \micron$.}
\tablenotetext{b}{IRAS03282+3035.}
\end{table}
\end{center}

\begin{center} 
\begin{table}
\caption{Sensitivity of Model Parameters\tablenotemark{a}\label{phystest}}
\begin{tabular}{lcrc}
\tableline\tableline
Variable 	& Range\tablenotemark{b}	
				& $\Delta p$ 	& $\Delta$\lint	\\
		& 		& 		& (\lsun )	\\
\tableline
$r_i$		& $30$ -- $240$ AU  & $< 0.1$	& $0$		\\
$r_o$		& $30,000$ -- $120,000$ AU & $< 0.1$	& $0.2$		\\
$n_f$		& $5 \times 10^5$ -- $3 \times 10^6$ \cmv & $< 0.1$	& $0.2$		\\
\lint		& $1.5$ -- $3.0$ \lsun		& $< 0.1$	& ...		\\ 
$\kappa _{\nu}$ & OH5 vs. OH2 	& $< 0.1$	& $1.9$		\\
\sisrf		& $0.3$ -- $3.0$ & $\pm 0.2$	& $1.5$		\\
$P_n(\theta )$\tablenotemark{c} & Jan. vs. Apr. & $\pm 0.1$		& $0$		\\
$p$		& $1.5 - 2.5$	& ...		& $0.3$		\\
\tableline
\end{tabular}
\tablenotetext{a}{Tested on B335.}
\tablenotetext{b}{The range of the variable tested.}
\tablenotetext{c}{The beam shape.}
\end{table}
\end{center}

\begin{center} 
\begin{table}
\caption{Shu77 Models \label{shumods}}
\begin{tabular}{llllllllllll}
\tableline\tableline
Source  & $r_{infall}$ & $a_{eff}$  & $r_i$ & $r_o$   & $\kappa _\nu$ &
$L_{int}$ & $I_{\nu}^{ISRF}$ & $\chi ^2_{850}$ & $\chi ^2_{450}$ & $\chi ^2_{SED}$ & Notes \\
& (AU)       & (km/s) & ($\cmv$)  & (AU)  & (AU)    &               &
(\lsun )&                  &                     &                 &               \\ 
\tableline
B335    & \\
& 6200    & 0.23  & 60    & 60,000  &  OH5 &  6.5 & 1.0 & 79.0  & 53.7  & 108  \tablenotemark{a} &      \\ 
& 6200    & 0.26  & 60    & 60,000  &  OH5 &  5.0 & 1.0 & 82.1  & 52.8  & 62.2 \\
& 1000    & 0.23  & 60    & 60,000  &  OH5 &  3.8 & 1.0 & 7.6   & 3.6   & 36.7        \\
& 500	  & 0.23  & 60    & 60,000  &  OH5 &  3.0 & 1.0 & 9.0   & 3.5   & 15.8        \\ 
& 1000	  & 0.23  & 60    & 60,000  &  OH5 &  4.5 & 0.3 & 2.3   & 0.8   & 66.8 \tablenotemark{a} & Best $\chi^2_r$   \\ 
& 1000	  & 0.26  & 60    & 60,000  &  OH5 &  4.0 & 0.3 & 2.4   & 1.0   & 39.1 \\
& 6200	  & 0.23  & 60    & 60,000  &  OH2 &  6.5 & 0.3 & 39.6  & 21.9  & 59.0 \tablenotemark{a} \\
& 1000    & 0.23  & 60    & 60,000  &  OH2 &  4.7 & 0.3 & 2.6   & 1.1   & 58.8 \tablenotemark{a} \\
B228    & \\ 
& 1000    & 0.23  & 60    & 30,000  &  OH5 &  1.0 & 0.3 & 2.9   & 15.6  & 32.2 & Best $\chi^2_{tot}$ \\ 
L723    & \\ 
& 1000    & 0.29  & 60    & 60,000  &  OH5 &  2.6 & 0.3 & 1.1   & 0.6   & 2.49 & Best $\chi^2_{tot}$ \\ 
IRAS03282\tablenotemark{b} \\
&  1000  & 0.23  & 60    & 45,000  &  OH5 & 1.0  & 0.3 & 4.8   & 13.3  & 34.1 \tablenotemark{a}       \\
&  1000  & 0.26  & 60    & 45,000  &  OH5 & 1.0  & 0.3 & 3.9   & 9.5   & 18.7 & Best $\chi^2_{tot}$ \\
L1448C  \\
&  1500  & 0.38  & 60    & 45,000  &  OH5 & 5.9  & 1.0 & 10.5  & 3.0   & 2.4  & \\
\tableline
\end{tabular}
\tablenotetext{a}{Unable to simultaneously match $S_{850}$ and $L_{bol}$.}
\tablenotetext{b}{IRAS03282+3035.}
\end{table}
\end{center}

\begin{center} 
\begin{table}
\caption{Properties of Best Fit Models \label{bestmods}}
\begin{tabular}{lllllllllll}
\tableline\tableline
Source  & $p$   & $n_f$             & $L_{obs}$ & $L_{int}$  & $L_{bol}^{mod}$  
& $\alpha_{450/850}^{120}$ & $M_{env}^{120}$ & Aspect\tablenotemark{a} & $T_{iso}$  & $M_{vir}^{120}$ \\
    &    &    &    & &  & & & Ratio & & \\
        &         & (\cmv )           & (\lsun ) & (\lsun )   & (\lsun ) 
&                          & (\msun )        &      & (K) & \msun \\
\tableline
B335    & 1.8     & $1.5 \times 10^6$ & 3.1(0.1) & 3.3  & 3.1  & 2.5 & 2.6 & 1.08 & 13.2  &  4.8\\
B228    & 1.9     & $1.2 \times 10^6$ & 1.2(0.2) & 1.0  & 1.1  & 2.6 & 0.8 & 1.22 & 12.5  &  2.8\\
L723    & 1.8     & $1.8 \times 10^6$ & 3.3(0.2) & 2.6  & 3.4  & 2.5 & 3.9 & 1.42 & 12.4  &  7.7\\ 
IRAS03282\tablenotemark{b} & 1.9 & $1.9 \times 10^6$   & 1.2(0.3) & 1.0  & 1.2  & 2.4 & 2.4 & 1.03 & 11.5  &  3.6\\
L1448C\tablenotemark{c}  & 1.7     & $2.6 \times 10^6$ & 6.0(0.5) & 5.9  & 6.0  & ... & 4.5 & ...  & ...   &  ...\\
L1527   & 1.1     & $6.5 \times 10^5$ & 2.2(0.2) & 2.2  & 2.2  & 2.4 & 1.6 & 1.59 & 15.0  &  2.6\\
L483    & 1.2     & $6.0 \times 10^5$ & 13(2) & 13.0 & 12.9 & 2.7 & 2.3 & 1.93 & 18.0  &  5.4\\
\tableline
\end{tabular}
\tablenotetext{a}{Ratio of major to minor axis for the 20\% peak contour.}
\tablenotetext{b}{IRAS03282+3035.}
\tablenotetext{c}{Quantities calculated using a 120\as\ aperture not shown due to
confusion from nearby sources.}
\end{table}
\end{center}

\end{document}